\documentclass[preprint,amsmath,amssymb,aps,
pra,superscriptaddress]{revtex4-1}
\usepackage{graphicx}
\usepackage{dcolumn}
\usepackage{bm}
\usepackage[usenames, dvipsnames]{color}

\DeclareUnicodeCharacter{0301}{\'{i}}
\begin{document}

\title{Vortices in a Ginzburg Landau Theory of Superconductors with Nematic Order}
\author{R. S. Severino}
\affiliation{Departamento de F\'{i}sica, FCEyN, Universidad de Buenos Aires,
and IFIBA, CONICET-Universidad de Buenos Aires, Ciudad Universitaria, CP 1428,   Buenos Aires, Argentina}
\author{P. D. Mininni}
\affiliation{Departamento de F\'{i}sica, FCEyN, Universidad de Buenos Aires,
and IFIBA, CONICET-Universidad de Buenos Aires, Ciudad Universitaria, CP 1428,   Buenos Aires, Argentina}
\author{E. Fradkin}
\affiliation{Department of Physics and Institute for Condensed Matter Theory, University of Illinois  at Urbana-Champaign,
1110 West Green Street, Urbana, Illinois 61801-3080, USA}
\author{V. Bekeris}
\affiliation{Departamento de Física, FCEyN, Universidad de Buenos Aires,
and IFIBA, CONICET-Universidad de Buenos Aires, Ciudad Universitaria, CP 1428,   Buenos Aires, Argentina}
\author{G. Pasquini}
\affiliation{Departamento de Física, FCEyN, Universidad de Buenos Aires,
and IFIBA, CONICET-Universidad de Buenos Aires, Ciudad Universitaria, CP 1428,   Buenos Aires, Argentina}
\author{G. S. Lozano}
\affiliation{Departamento de Física, FCEyN, Universidad de Buenos Aires,
and IFIBA, CONICET-Universidad de Buenos Aires, Ciudad Universitaria, CP 1428,   Buenos Aires, Argentina}
\begin{abstract}

In this work we explore the interplay between superconductivity and nematicity in the framework of a Ginzburg Landau theory with a nematic order parameter coupled to the superconductor order parameter, often used in the description of superconductivity of Fe based materials. In particular, we focus on the study of the vortex-vortex interaction  in order to determine the way nematicity affects its attractive or repulsive character. To do so, we use a dynamical method based on the solutions of the Time Dependent Ginzburg Landau equations in a bulk superconductor. An important contribution of our work is the implementation of a pseudo-spectral method to solve the dynamics, known to be highly efficient and of very high order in comparison to the usual finite differences/elements methods.  The coupling between the superconductor and the (real) nematic order parameters is represented by two terms in the free energy: a biquadratic term and a coupling of the nematic order parameter to the covariant derivatives of the superconductor order parameter. Our results show  that there is a competing effect: while the former independently of  its competitive or cooperative character generates an attractive vortex-vortex interaction, the latter  always   generates a repulsive interaction. 
\end{abstract}

\maketitle

\section{Introduction}

The nature of the vortex-vortex interaction in Ginzburg Landau (GL) type theories has attracted much attention over the years, both from the condensed matter community, where GL theories are the succesful phenomenological theory of BCS superconductors \cite{GL, abrikosov}, and also from the High Energy community where vortices appear as non pertubative solutions of the Higgs model and other quantum field theories \cite{Nielsen}.

An important first step in the description of the interactions was taken by Kramer \cite{kramer},
who gave a simple (approximate) expression for the vortex-vortex interaction, showing that vortices repel for $\kappa>1/\sqrt{2}$ and attract for $\kappa<1/\sqrt{2}$, where $\kappa=\lambda_L/\xi$ is the GL parameter defined as the ratio of the London magnetic penetration  length and the superconductor coherence length. The particular critical value $\kappa_c=1/\sqrt{2}$ also signals the boundary between Type I and Type II superconductors characterized by the presence of an Abrikosov vortex lattice phase. This particular value of $\kappa_c$ holds for standard macroscopic 3-dimensional superconductors, and the problem has been revisited by many authors using a variety of techniques and applied to different systems of interest \cite{jacobs,brandt,speight1,mackensie,bettencourt,mohamed,hernandez,auzzi,babaev}. A very detailed numerical analysis of the vortex-vortex interaction in the context of the standard GL model has been performed by Chaves et al.~\cite{chaves}

The non interacting character of vortices for $\kappa_c$ in the GL model can be established analytically using the Bogomol'nyi identity and showing that  the energy per unit length is proportional to the vorticity (or the number of vortices) \cite{bogomolny}.
From a mathematical point of view the value $\kappa = 1/\sqrt{2}$, commonly referred to as the \textit{self dual point}, is very interesting. It can be shown that the second order Euler Lagrange equations are in this case equivalent to a set of much simpler first order equations known as self-dual equations. In High Energy Physics, where the Ginzburg Landau free energy corresponds to the static energy of the $U(1)$ Higgs model, these equations are known  under the name of Bogomol'nyi-Prassad-Sommerfeld (BPS) and were studied originally in Refs  \cite{bogomolny, schaposnik} for the case of vortices, and  for the case of monopoles in Ref.~\cite{prasad}. In the Superconductor literature these equations (in an axially symmetric ansatz) were first discussed in Ref.~\cite{harden} 

The existence of static configurations can be proved rigorously and the space of solutions can be fully characterised (\textit{moduli space}) not only for the geometry of the plane (which we are dealing with in this work) but also for more general geometries (in general manifolds without boundaries). Furthermore, the existence of a self-dual point also indicates the presence of more \textit{exotic} symmetries which in addition to the existing fields
of the theory (represented by standard commuting fields) involve additional fields that are represented by anticommuting Grassman
variables (\textit{supersymmetries}) (see for instance \cite{shifman_yung_2009} and references therein). These supersymmetries play a fundamental role in the understanding of the non perturbative sector of modern quantum field theories. Self dual points exist for many theories with multiple order parameters but not for all of them. When
self dual equations do not exist, the nature of vortex-vortex interaction must be studied numerically.

In this work we are interested in Ginzburg Landau theories with nematic order parameters. The existence of an electronic nematic phase in strongly correlated systems has been originally theoretically proposed in Refs.~\cite{kivelson1, kivelson2}, and a large amount of experimental evidence has been documented during the last decade. In particular, an anisotropic phase has been reported in the underdoped regime of both cuprate  \cite{lavrov,keimer,taillefer,damascelli} and Fe-based \cite{dai,fisher,prozorov1,prozorov2} high-temperature superconductors with a concurrent breaking of the $C_4$ symmetry in the structural and transport properties, driven by electronic degrees of freedom \cite{davis,prozorov1}. In this framework, the role played by nematicity in unconventional superconductivity has been theoretically explored in terms of coupling between the nematic and superconducting order parameters \cite{kivelson3, kivelson4}. Growing experimental evidence points towards this connection,
where competition and cooperation are suggested in iron based superconductors BaFeAs \cite{Nandi, Kalisky1}  and FeSe \cite{song}, respectively \cite{schmalian}. However, the  character of this coupling  is still matter of debate \cite{schmidt,Chen}.   

In particular, we are interested in exploring how the presence of the nematic order alters the vortex-vortex interaction. We focus on how the new parameter affects the boundary between type I and type II superconductivity in bulk samples (by bulk samples we mean 3-dimensional materials where any characteristic lengths such as the London penetration length or the superconductor coherence length are much smaller than the dimension of the sample; in mesoscopic samples and/or films the phenomenology is expected to be different, see for example Ref.~\cite{deo} for standard BCS superconductors). We will also show that if the nematic order parameter is taken constant in space and time as an external field, the existence of a self dual point can be easily established.

In the more general case, we will study the vortex-vortex interaction numerically. In order to do so, we will use a {\em dynamical} technique. That is, we will address the problem of vortex-vortex interaction by studying the dissipative dynamics as dictated by the time dependent Ginzburg Landau (TDGL) equation \cite{schmid}. The TDGL model has a very long history, and it can be used for a variety of purposes \cite{aranson}. 

Contrary to other approaches that require the use of approximations or particular ansatze, a dynamical approach gives a direct access to the characterization of the vortex-vortex interaction. A very relevant point of our work is the particular numerical method that we apply for solving the TDGL that gives us a high level of control over the sources of numerical uncertainties. The most extended technique to solve this equations in studies of superconductivity is via the method of finite differences. Although it has been successfully  applied in many cases (for an example see \cite{sadovsky}), from a practical point of view finite difference methods are resource expensive and in some cases they might even become unstable. Moreover, they display numerical dissipation and numerical dispersion that can result in spurious solutions especially when implemented at the lowest orders in the derivatives \cite{boyd2}. By numerical dissipation we mean an enhanced dissipation which appears as the result of the implementation of a numerical technique
that might result in certain cases in larger than expected damping of high-frequency modes when performing a numerical integration. In other words, an artificial viscosity of numerical origin. Numerical dispersion is a spurious dispersivity also resulting from errors in the numerical method; for example,  in a weakly interacting bosonic superfluid, it manifests as a numerical dispersion relation that is more dispersive than the expected Bogoliuvov prediction, or in other words, in spatial modes that propagate faster than expected. This effect shows up in the simulations as spurious moving oscillations or wiggles in the fields \cite{ghosta, ghostb, gpesolver}.

We use here {\em pseudo-spectral methods} instead. These methods have been applied in many areas, but most importantly they have been recently used and optimized for the study of turbulence in quantum fluids (i.e., the disorganized spatio-temporal evolution of quantized vortices in superfluids and Bose-Einstein condensates) a domain where they have been applied very successfully \cite{ghosta, ghostb, gpesolver}. The problem of quantum fluids is very close to that of superconductors as in that field TDGL dynamics serve a first step in the preparation of initial conditions. So, the application of these methods to superconductors can be done with a simple adaptations of codes already developed in Refs.~\cite{ghosta,ghostb, gpesolver}, and are expected to give excellent results concerning precision and performance. In particular, pseudo-spectral methods have no numerical dispersion nor dissipation, and thus (when properly implemented and at the proper spatial resolution) they only present the dissipation that naturally arises from the nature of the equations of motion. They can also reproduce the exact dispersion relation of the physical system without numerical contamination. All these properties result from the fact that these methods display exponentially rapid convergence of the numerical solutions with the increase of spatial resolution (against algebraic convergence in finite differences methods, see Ref.~\cite{Boyd} for a detailed comparison).

Our work is organized as follows. In section II we introduce the free energy and the TDGL equations describing our model. In section III we introduce the numerical method and we apply it to the standard GL model (without nematicity), with the main purpose of validating the method against known results. In section IV we apply the method to the specific problem of vortex-vortex interacions for the model with nematicty, and we leave for section V a list of the main results of our work together with some discussions concerning future lines of research. 
 
\scriptsize
\section{Ginzburg-Landau model with nematic order parameter and TDGL equations}
\normalsize

The Hemholtz free energy of the original Ginzburg Landau (GL) model can be written as 
\begin{equation}
    F_s= \int dV \left[ \alpha_{GL} |\psi|^2 +\frac{\beta_{GL}}{2} |\psi|^4 + \frac{\hbar^2}{2m} \left|\boldsymbol{\mathcal{D}}\psi\right|^2 + \frac{(\nabla \times \boldsymbol{A})^2}{8\pi}\right] ,
    \label{eq:energia_compacta}
\end{equation}
where $\psi$ is the complex superconducting order parameter related to the superficial density via $|\psi|^2=n_s$, $\boldsymbol{A}$ is the vector potential related to the magnetic induction as $\nabla\times\boldsymbol{A}=\boldsymbol{B}$, and $\boldsymbol{\mathcal{D}}=-i\nabla-\frac{e}{\hbar c}\boldsymbol{A}$ is the covariant derivative. The fields have units of $[\psi] = \frac{1}{[L]^{3/2}}$ and $[A]=\frac{[M][L]^2}{[Q][T]}$ (where $[L],[M],[Q]$, and $[T]$ stand respectively for length, mass, charge, and time units). The parameters $ \alpha_{GL}$ and $\beta_{GL}$  depend on the temperature, more specifically, $\alpha_{GL}=\alpha_0(T-T_c)$, changing sign at $T_c$ signalling the transition to the superconducting phase. Here $m$ is a parameter with dimensions of mass but notice that it is not directly linked to the mass of any particular particle. It can be linked to the phase stiffness of the order parameter. We will keep this notation as it is the most widely used in the literature. The charge of the Cooper-pairs (twice the electron charge) will be noted as $e$ while $\hbar$ and $c$ are the Planck constant and the speed of light respectively. Here, and for the rest of our work,  we consider vortex like solutions that are translational invariant along the $\hat{z}$-direction (with the magnetic field pointing in the $+\hat{z}$-direction), so any $z$ dependence of fields will be ignored. In writing Eq.~(\ref{eq:energia_compacta}) we have also assumed that the superconductor is isotropic in the $x-y$ plane, this fact is not strictly true in Fe-based superconductors (this will have important consequences for the character of the nematic order parameter as we discuss  bellow).

The superconducting current is given by:
\begin{equation}
    \boldsymbol{j}=\frac{\hbar e}{2im}(\psi^{*}\nabla\psi-\psi\nabla\psi^{*})-\frac{e^2}{mc}\boldsymbol{A}|\psi|^2 .
     \label{current_gleq}
\end{equation}

A time dependent modification of the GL equations can be established under the assumption that the derivative of the free energy is a generalized force. Energy dissipation can happen in the system either in the form of heat (related to the Joule effect due to the normal cores of the vortices) and/or through irreversible variation of the order parameter \cite{schmid}. Thus, the purely dissipative dynamics of the model is given by the equations 
\begin{equation}
    \frac{\hbar^2}{2mD} \partial_t \psi=-\frac{\delta F_s}{\delta \psi^{*}}
    \quad \quad \frac{\sigma}{c^2}\partial_t \boldsymbol{A}= -\frac{\delta F_s}{\delta \boldsymbol{A}} .
    \label{eq:dyn_a}
\end{equation}
Here $\sigma$ is the electrical conductivity with units of $[\sigma] = [T]^{-1}$, and $D$ is a diffusion constant  with units of $[D] =[ L]^2 [T]^{-1}$. Notice that we are neglecting a term proportional to the second time derivative of ${\boldsymbol{A}}$ and that we are working in the gauge $A_0=0$. Equations (\ref{eq:dyn_a}) are known as the Time Dependent Ginzburg Landau (TDGL) equations and were introduced more than 50 years ago by Schmid \cite{schmid}. A stochastic noise term related to thermal fluctuations can be added, and then they become Langevin-type evolution equations, suited for superconductors 

Note that the numerical method we will introduce next is specially well suited for this approach, as the pseudo-spectral numerical truncation preserves the Langevin structure of the equations (see \cite{shukla} for a discussion of thermal fluctuations in the context of superfluids using this method). However, in this paper we will only take account of the temperature via the standard temperature dependence of the parameters of the free energy. We stress here that the dynamics are dissipative and, in the present case, this is a totally reasonable assumption. 

The dissipative character of the dynamics can be easily shown by considering that
\begin{eqnarray}
    \frac{dF_s}{dt}&=&\frac{\delta F_s}{\delta \psi^{*}}
    \partial_t \psi^* +\frac{\delta F_s}{\delta \boldsymbol{A}}\partial_t \boldsymbol{A}   =-\left(\frac{2mD}{\hbar^2}\frac{\delta F_s}{\delta \psi^{*}}\frac{\delta F_s}{\delta \psi}
     +\frac{c^2}{\sigma}\frac{\delta F_s}{\delta \boldsymbol{A}} \frac{\delta F_s}{\delta \boldsymbol{A}}\right) <0 .
\end{eqnarray}

In order to account for the nematic phase of Fe-based superconductors, an additional order parameter needs to be included in the free energy.  Due to the presence of the tetragonal and orthorrombic crystallographic structure of Fe-based superconductors, we will take the nematic order parameter to be a {\em real} field $\eta$, with the symmetry property that $\eta \rightarrow - \eta$ under a 90 degree rotation of the crystallographic structure. This should be contrasted for instance with the order parameter of nematic liquids, which involves a continuous rotation. The new terms in the Helmholtz free energy involving this  parameter can be written as:
\begin{equation}
      F_N = \int_V dV \left[\gamma_2(\nabla\eta)^2 +\gamma_3\eta^2  + \frac{\gamma_4}{2} \eta^4 +\frac{\hbar}{2m}\lambda_1 \eta(|\mathcal{D}_x\psi|^2-|\mathcal{D}_y \psi|^2) + \lambda_2 \eta^2\psi^2\right] .
    \label{eq:nem_energy}
\end{equation}
 
The first three terms correspond to the nematic free energy while the last two terms couple the nematic order parameter to the superconducting order parameter and (via the covariant derivative) to the vector field. While the biquadratic term does not depend on the nematic character of the order parameter, and it is in fact quite common in several theories with multiple order parameters (for an example on multiband superconductors see Ref. \cite{babaev_83}), the  term proportional to $\lambda_1$  depends specifically on the nematic nature of $\eta$, since it would not be present otherwise as $\eta \longrightarrow -\eta$ when we interchange $x \longleftrightarrow y$. GL theories of this type were considered for instance in Ref.~\cite{chowdhuri}  to study vortices is FeSe compounds (notice though that they work in the limit $\lambda_L \rightarrow \infty$, that amounts to neglecting gauge field dynamics and structure). Vortices were also considered in Refs.~\cite{oblicuos, putilov} (for a case in which the nematic parameter is complex see \cite{barci}). Superconducting-nematic coupling of this kind was also considered in \cite{schmidt} in the study of strain detwined mixed states \cite{sanches_nature}. In a more general setting additional terms incorporating strain, stress, and their couplings to the order parameters can also be included \cite{strain1,strain2,strain3,strain4,strain5}.

The dynamics of the nematic order parameter is  prescribed by:
\begin{equation}
    \frac{\hbar^2}{2mD_{n}}\partial_t \eta =- \frac{\delta F}{\delta \eta} ,
\end{equation}
where $D_n$ is the nematic difussion constant. 

All parameters depend in principle on the temperature. The thermodynamical  phases of the model are determined by the signs of $\alpha_{GL}$ and $\gamma_3$, and by the sign and value of $\lambda_2$. We will assume that we are in a case where {\em both} symmetries are broken for every value of $\lambda_2$. This implies that $\alpha_{GL}<0$ and $\gamma_3<0$. Further restrictions on $\lambda_2$ will be soon derived.

Important parameters of our model are the different lengths associated to each order parameter
\begin{equation}
    \xi^2 = \frac{\hbar^2}{2m|\alpha_{GL}|}, \quad\quad
    \lambda_L^2 =\frac{mc^2}{4\pi e^2 \rho_0}, \quad\quad
    l_\eta^2=\frac{\gamma_2}{|\gamma_3|},
\end{equation}
where $\xi$ is the superconductor coherence length, $ \lambda_L$ is the London length, and $l_\eta$ is the nematic coherence length $\left( \text{here} \hspace{1mm} \rho_0=\frac{|\alpha_{GL}|}{\beta_{GL}}\hspace{1mm}\text{with} \hspace{1mm}[\rho_0] = [L]^{-3}\right)$. Next, we rescale the vector field and the order parameters as 
\begin{equation}
    \psi = \sqrt{\rho_0}\tilde{\psi}, \quad\quad \boldsymbol{A} = \frac{mc |\alpha_{GL}|}{e\hbar}\boldsymbol{a}, \quad\quad \eta = \eta_0 \tilde{\eta},
\end{equation}
where $\eta_0^2 = -\frac{\gamma_3}{\gamma_4}$ . Note that this redefinition implies that the magnetic vector potential has units of $[\boldsymbol{a}] = [L]$. 
Next we define new coefficients for the nematic free energy in eq.~(\ref{eq:nem_energy}) as
\begin{equation}
     \hat{\lambda}_1 = \frac{\lambda_1\eta_0}{\hbar} \quad;\quad \hat{\lambda}_2 = \frac{\lambda_2\eta_0^2}{|\alpha_{GL}|}, \quad\quad
  \Gamma_2 = \frac{\gamma_2 }{|\alpha_{GL}|(\rho_0/\eta_0^2)}, \quad\quad \Gamma_4 = \frac{\gamma_4\eta_0^2}{(|\alpha_{GL}|\rho_0/\eta_0^2)}.
\end{equation}
Notice that $\hat{\lambda}_1, \hat{\lambda}_2$, and $\Gamma_4$ are dimensionless, and $[\Gamma_2]=[L]^2$. Finally, the (dimensionless) GL parameter is
\begin{equation}
     \kappa = \frac{\lambda_L}{\xi}.
    \label{eq:glparameter}
\end{equation}

Using the newly defined parameters and fields, the free energy can be expressed as:
\begin{eqnarray}
     F = |\alpha_{GL}| \rho_0 \int dV \left[ \frac{1}{2}(|\tilde{\psi}|^2-1)^2+\xi^2 |\nabla\tilde{\psi}|^2 - \boldsymbol{a}\text{Im}(\tilde{\psi}^{*}\nabla\tilde{\psi})+\frac{1}{4\xi^2}\boldsymbol{a}^2\tilde{\psi}^2+ \right.\\ \nonumber
    \left.
     \frac{\kappa^2}{4} (\nabla\times\boldsymbol{a})^2+\Gamma_2(\nabla\tilde{\eta})^2
    +\frac{\Gamma_4}{2}(\tilde{\eta}^2-1)^2+\xi^2\hat{\lambda}_1\tilde{\eta}(|\mathcal{D}_x\tilde{\psi}|^2-|\mathcal{D}_y\tilde{\psi}|^2)+ \hat{\lambda}_2 \tilde{\eta}^2\tilde{\psi}^2\right],
\end{eqnarray}
where the rescaled covariant derivative is $\boldsymbol{\mathcal{D}}=-i\nabla-\frac{\boldsymbol{a}}{2\xi^2}$. It is possible to rewrite the theory  if we redefine the time variable in terms of a dimensionless parameter $\tau$,
\begin{equation}
    t = \frac{\hbar^2}{2mD|\alpha_{GL}|}\tau,
\end{equation}
and we rescale the electrical conductivity and the diffusion constant as
\begin{equation}
 \sigma_1 = \frac{4\pi\sigma}{c^2}\frac{2mD|\alpha_{GL}|}{\hbar^2}, \quad\quad D_\eta = D_n (\rho_0/\eta_0^2).
\end{equation}
This redefinition implies that $[\sigma_1]=[L]^{-2}$. The dynamics of the new fields are prescribed by:
\begin{equation}
    \partial_\tau \tilde{\psi} = -\left(\frac{1}{|\alpha_{GL}|\rho_0}\right)\frac{\delta F}{\delta \tilde{\psi}^{*}},
    \label{eq:dyna_psi}
\end{equation}
\begin{equation}
    \partial_\tau \boldsymbol{a} = -\left(\frac{1}{|\alpha_{GL}|\rho_0}\right)\frac{2}{\kappa^2 \sigma_1} \frac{\delta F}{\delta \boldsymbol{a}},
    \label{eq:dyna_a}
\end{equation}
\begin{equation}
     \frac{D}{D_\eta}\partial_\tau \tilde{\eta} =-\left(\frac{1}{|\alpha_{GL}|\rho_0}\right)\frac{\delta F}{\delta \tilde{\eta}},
\end{equation}
A straightforward calculation yields the following equation of motion for the order parameter:
\begin{gather}
\partial_\tau \tilde{\psi} = \xi^2\nabla^2\tilde{\psi} + \tilde{\psi}(1-|\tilde{\psi}|^2) -i\boldsymbol{a} \cdot \nabla\tilde{\psi}- \frac{i}{2}\tilde{\psi}\nabla \cdot \boldsymbol{a} - \frac{1}{4\xi^2}a^{2}\tilde{\psi} - \hat{\lambda}_2 \tilde{\eta}^2 \tilde{\psi} \\
\nonumber  - \frac{i \hat{\lambda}_1}{2}\tilde{\eta}(a_x\partial_x\tilde{\psi}-a_y\partial_y\tilde{\psi}) - \frac{\hat{\lambda}_1}{4\xi^2}\tilde{\eta}\tilde{\psi}(a_x^2-a_y^2) 
+ \xi^2 \hat{\lambda}_1(\partial_x(\tilde{\eta}\partial_x\tilde{\psi})-\partial_y(\tilde{\eta}\partial_y\tilde{\psi)}) \\
\nonumber -\frac{i\hat{\lambda}_1}{2}(\partial_x(a_x\tilde{\eta} \tilde{\psi})-\partial_y(a_y\tilde{\eta} \tilde{\psi}))
\label{eq:psi_din}
\end{gather}
For the nematic order parameter, we have 
\begin{gather}
\frac{D}{D_\eta}\partial_\tau\tilde{\eta} = 2\Gamma_2\nabla^2\tilde{\eta} + 2\Gamma_4\tilde{\eta}(1-\tilde{\eta}^2) - 2\hat{\lambda}_2\tilde{\psi}^2\tilde{\eta} - \xi^2\hat{\lambda}_1(|\partial_x\tilde{\psi}|^2-|\partial_y\tilde{\psi}|^2) \\ \nonumber   -\frac{\hat{\lambda}_1}{4\xi^2}\tilde{\psi}^2(a_x^{2} -a_y^{2}) 
+\hat{\lambda}_1(a_x\text{Im}(\tilde{\psi^{*}}\nabla_x\tilde{\psi})-a_y\text{Im}(\tilde{\psi^{*}}\nabla_y\tilde{\psi}))
\end{gather}
And for the components of the vector potential we find
\begin{eqnarray}
\partial_\tau {a}_x &=& \frac{2\left(1+\hat{\lambda}_1\tilde{\eta}\right)}{\kappa^2\sigma_1}\text{Im}(\tilde{\psi}^{*}\nabla_x\tilde{\psi}) - \frac{\left(1+\hat{\lambda}_1\tilde{\eta}\right)}{\kappa^2\xi^2\sigma_1}{a}_x|\tilde{\psi}|^2 - \frac{1}{\sigma_1}(\nabla\times\nabla\times\boldsymbol{a})_x , \\
\partial_\tau {a}_y &=& \frac{2\left(1-\hat{\lambda}_1\tilde{\eta}\right)}{\kappa^2\sigma_1}\text{Im}(\tilde{\psi}^{*}\nabla_y\tilde{\psi}) - \frac{\left(1-\hat{\lambda}_1\tilde{\eta}\right)}{\kappa^2\xi^2\sigma_1}{a}_y|\tilde{\psi}|^2 - \frac{1}{\sigma_1}(\nabla\times\nabla\times\boldsymbol{a})_y .
\end{eqnarray}

In order to find constraints for the parameters of our theory, we analyze the potential part of the Helmholtz free energy:
\begin{gather}
    V(|\tilde{\psi}|^2,\tilde{\eta}^2) = \rho_0 |\alpha_{GL}|  \int dV \left[\frac{1}{2}(|\tilde{\psi}|^2-1)^2 + \hat{\lambda}_2 \tilde{\eta}^2\tilde{\psi}^2 
    +\frac{\Gamma_4}{2}(\tilde{\eta}^2-1)\right] .
    \label{eq:potencial}
\end{gather}
The quartic terms in the previous expression constitute a quadratic form in $|\tilde{\psi}|^2$ and $\tilde{\eta}^2$, represented by the matrix
\begin{equation}
    \Gamma=  \frac{|\alpha_{GL}|\rho_0}{2}\begin{pmatrix}
    1 & \hat{\lambda}_2 \\
    \hat{\lambda}_2 & \Gamma_4
    \end{pmatrix} .
\end{equation}
Imposing that this form is positive-definite implies that:
\begin{equation}
    \hat{\lambda}_2 > -\sqrt{\Gamma_4} .
    \label{eq:cond0}
\end{equation}
Our theory also supposes that both the nematic and superconducting symmetries are broken in the initial state. We will consider $\tilde{\eta}_v$ and $\tilde{\psi}_v$ to be the equilibrium values of the nematic and superconducting order parameters and the minima for the potential defined in Eq.~(\ref{eq:potencial}). These values must satisfy that $\nabla V (\tilde{\eta}_v,\tilde{\psi}_v) = 0$. This relation defines a linear system of equations for $\tilde{\eta}_v$ and $\tilde{\psi}_v$, given by
\begin{equation}
 |\tilde{\psi}_v|^2+\hat{\lambda}_2\tilde{\eta}_v^2 = 1 , \quad\quad \frac{\hat{\lambda}_2}{\Gamma_4}|\tilde{\psi}_v|^2+\tilde{\eta}_v^2 = 1 .
\end{equation}
Solving the system yields the equilibrium values of the uniform state (the minima of the potential defined in Eq.~\ref{eq:potencial}) for each order parameter
\begin{equation}
    |\tilde{\psi}_v|^2 = \frac{1-\hat{\lambda}_2}{1-(\hat{\lambda}_2^2/\Gamma_4)} , 
    \quad
    \tilde{\eta}_v^2 = \frac{1-(\hat{\lambda}_2/\Gamma_4)}{1-(\hat{\lambda}_2^2/\Gamma_4)} .
    \label{eq:eta_vacuum}
\end{equation}
These minima exist if the constants of the model satisfy certain relations that will determine the range of values the parameters of the theory can take. For the biquadratic coupling parameter $\hat{\lambda}_2$, we have that

\begin{equation}
    \hat{\lambda}_2 < 1 ,  \quad\quad
    \hat{\lambda}_2 < \Gamma_4 ,  \quad\quad |\hat{\lambda}_2| < \sqrt{\Gamma_4} .
\end{equation}
Therefore, the admitted values for the biquadratic coupling between the superconductor and nematic order parameters are
\begin{equation}
-\sqrt{\Gamma_4}   < \hat{\lambda}_2< \text{min} ( 1, \Gamma_4) .
\label{eq:range_gam1}
\end{equation}
Finally, from equation (\ref{eq:psi_din}), notice that the terms involving second derivatives of $\tilde{\psi}$ must be positive definite when $ \tilde{\eta} = \tilde{\eta}_v$. Imposing this condition also implies that the value of $\hat{\lambda}_1$ must be bounded for the problem to remain stable. For a constant nematic order parameter, the terms involving second derivatives of the superconducting order parameter are
\begin{equation}
    \xi^2 ( 1 + \hat{\lambda}_1 \tilde{\eta}_v ) \partial_x^2 \tilde{\psi} + \xi^2 ( 1 - \hat{\lambda}_1 \tilde{\eta}_v ) \partial_y^2 \tilde{\psi} ,
\end{equation}
which implies that:
\begin{equation}
    |\hat{\lambda}_1| < \frac{1}{\tilde{\eta}_v} < \sqrt{\frac{1-(\hat{\lambda}_2^2 / \Gamma_4)}{1-(\hat{\lambda}_2/\Gamma_4)}} .
\end{equation}

Having stated the theoretical bases that determine the free energy that describes our system, we will analyze in the next section the vortex-vortex interaction in the pure GL model (no nematic order)  in order to validate our numerical method against known results.

\section{Vortices in TDGL dynamics using pseudo-spectral methods}\label{sec:GL}

In this section we introduce the basic tools used for the numerical solution of the TDGL equations, while we refer the reader to the literature for more specific details on pseudo-spectral methods \cite{boyd2}. We show first how the method works in the standard Ginzburg Landau problem (without nematic order) and then how some known results can be recovered. The use of a dynamical method allows for the calculation of the energy and the intervortex distance as a function of time, a valuable asset that can be used to characterize the interaction force between vortices without extra assumptions. Moreover, it can become  useful in the study of  stable (and sometimes unstable) fixed points of the equations.

\subsection{General considerations and initial conditions}

The code used for the simulations is the Geophysical High-Order Suite for Turbulence, or GHOST for short \cite{ghosta, ghostb}, which is an accurate and highly scalable pseudo-spectral code that has been succesfully applied to solve a variety of Partial Differential Equations often encountered in studies of turbulent flows and in magnetohydrodinamics \cite{hdsolver,mhdsolver,mhdhallsolver}, and more recently in superfluids \cite{gpesolver}.

The pseudo-spectral method is based on a Fourier decomposition of each dynamical variable (both order parameters and the magnetic field) into a set of finite (but large) Fourier modes, and solves a system of equations that determines the time evolution of the Fourier coefficients of said decomposition. This ensures the exponentially fast convergence of the method. Computation of spatial derivatives turns into products in Fourier space, which can be computed efficiently. Nonlinear terms, on the other hand, become convolutions, whose computation in one-dimension require $\mathcal{O}(N^2)$ operations (where $N$ is the linear spatial resolution). To circumvent this cost, the Fourier transform of the variables is computed (as Fourier transforms require $\mathcal{O}(N \ln N)$ operations), and products of two fields are computed in real space, and dealiased after transforming back to Fourier space. It can be shown that this process is equivalent to performing an exact Galerkin truncation of the system \cite{boyd2}. As they are based on a Fourier basis, pseudo-spectral methods are known to be optimal on periodic domains, and thus fields are restricted to specific geometries and configurations. Here, to show the applicability of the method we consider this case, even though generalization to non-periodic domains is possible using other expansion basis \cite{boyd2}, or more generally using a Fourier continuation method \cite{Fontana_2020}. Such generalizations will be considered elsewhere. 

Our simulation box consists of a $2\pi L_x \times 2\pi L_y $ region in the $xy$ plane, with $L_x$ and $L_y$ parameters that can be chosen arbitrarily. The numerical method also allows for a third length $L_z$ to be chosen, but in our particular case this choice is irrelevant since we are dealing with translational invariance in the $\hat{z}$ direction and the relevant physics takes place in the $xy$ plane.  We fixed $L_x = L_y = L=1$ with a spatial resolution of $N = 512$ points in each direction, which in turn means that our simulation box is $2\pi$ long in each direction. Although we have chosen $L=1$,  we will keep writing it when necessary in order to keep track of the length dimensions. These choices define a uniform two-dimensional (2D) grid on the simulation box with spacing between points in each axis of $\Delta r = \frac{2\pi}{512} \cong 0.012 L$, which determines the scale of the smallest lengths we can resolve. We fixed the superconducting coherence length as $\xi = 0.04 L$, around three times the size of the spatial separation in each direction. This choice allows us to accurately resolve the vortex core, related to the mathematical singularity of the initial condition (see the discussion around Eq.~(\ref{eq:initialz}); note that a finite differences method would require a finer grid to achieve similar accuracy \cite{boyd2}).

The TDGL equations for the standard superconductor (i.e, with no coupling to a nematic order parameter) are then,
\begin{equation}
       \partial_\tau \tilde{\psi} = \xi^2\nabla^2\tilde{\psi} + \tilde{\psi}(1-|\tilde{\psi}|^2) -i\mathbf{a} \cdot \nabla\tilde{\psi}- \frac{i}{2}\tilde{\psi}\nabla \cdot \mathbf{a}-\frac{1}{4\xi^2}a^{2}\tilde{\psi} ,
\end{equation}
\begin{equation}
   \partial_\tau \mathbf{a} =\frac{2}{\kappa^2\sigma_1}\text{Im}(\tilde{\psi}^{*}\mathbf{\nabla}\tilde{\psi})-\frac{1}{\kappa^2\xi^2\sigma_1}\mathbf{a}|\tilde{\psi}|^2-\frac{1}{\sigma_1}(\nabla\times\nabla\times\mathbf{a}) .
   \label{eq:a_din}
\end{equation}
The first of these equations is the same as the one appearing in Ref.~\cite{nore} in the study of quantum turbulence flows. The main difference is that in that work the vector field $\mathbf{a}$ (which in that context is a velocity field)  is {\em fixed}, whereas here \textbf{a} is the magnetic vector potential and  it is {\em dynamical}.

As already mentioned, these equations are discretized in space using a Fourier expansion, and in time using a Runge-Kutta method of second order. Spatial derivatives are computed in Fourier space, while non-linear terms in the fields are computed in real space and dealiased using the 2/3 rule to truncate the resulting Fourier expansion and control aliasing instabilities \cite{shukla}. The $2/3$ rule for dealiasing is a filter in which all modes with wave number $k > k_{max} = N/3$ (where $N$ is the linear resolution) are set to zero. In other words, the Fourier series is truncated up to a maximum wavenumber $k_{max}$ which implies that, when compared with the maximum Nyquist frequency $k_N = N/2$, this truncation preserves $2/3$ of all modes in Fourier space. Therefore, dealiasing is just the elimination of aliasing in the product of Fourier-projected fields by filtering its components with the highest wave numbers. This, together with the condition $\xi > \Delta r$, ensures the exponentially fast spatial convergence of the solutions. Since the system is dissipative, the diffussion term (proportional to $\nabla^2$) is responsible for setting the time scale used for the numerical simulations. In order to find a limit to the time step necessary for convergence of the numerical method, we have to calculate the Courant number $C$ through the Courant-Friedrichs-Levy condition (CFL, for short). Conventionally, it is stated that $C < 1$ to guarantee convergence. We construct two Courant numbers, $C_{\tilde{\psi}}$ and $C_{a}$, through analyzing the diffusive terms in the equations. Through dimensional analysis, we can state that:
\begin{gather}
    C_{\tilde{\psi}} = \xi^2 \frac{\Delta \tau}{\Delta r^2} , \quad \quad C_{a} = \frac{1}{\sigma_1} \frac{\Delta \tau}{\Delta r^2} .
\end{gather}
For the range of parameters used in this paper, a time step of $\Delta \tau < 9\times 10^{-4}$ is enough to guarantee that both Courant numbers are within the desired range, thus guaranteeing convergence. 

For the initial condition on the superconducting order parameter we follow \cite{nore}, and start with  single vortex-like configuration in the $xy$ plane  of the form:
\begin{equation}
   \tilde{\psi} (x, y, t=0) = \tilde{\psi}_v \frac{(\lambda + i \mu)}{\sqrt{\lambda^2 + \mu^2}}\tanh{\left(\frac{\sqrt{\lambda^2 + \mu^2}}{\sqrt{2}\xi}\right)} ,
    \label{eq:initialz}
\end{equation}
where
\begin{equation}
 \lambda = \sqrt{2} \cos{(x)} ,  \quad \quad 
 \mu = \sqrt{2}\cos{(y)} ,
\end{equation}
which in fluid dynamics are related to the Clebsch potentials, and are chosen to ensure periodicity, required by the Fourier expansion used to solve the equations. 

\begin{figure}
\centering
\includegraphics[width=1.0\textwidth]{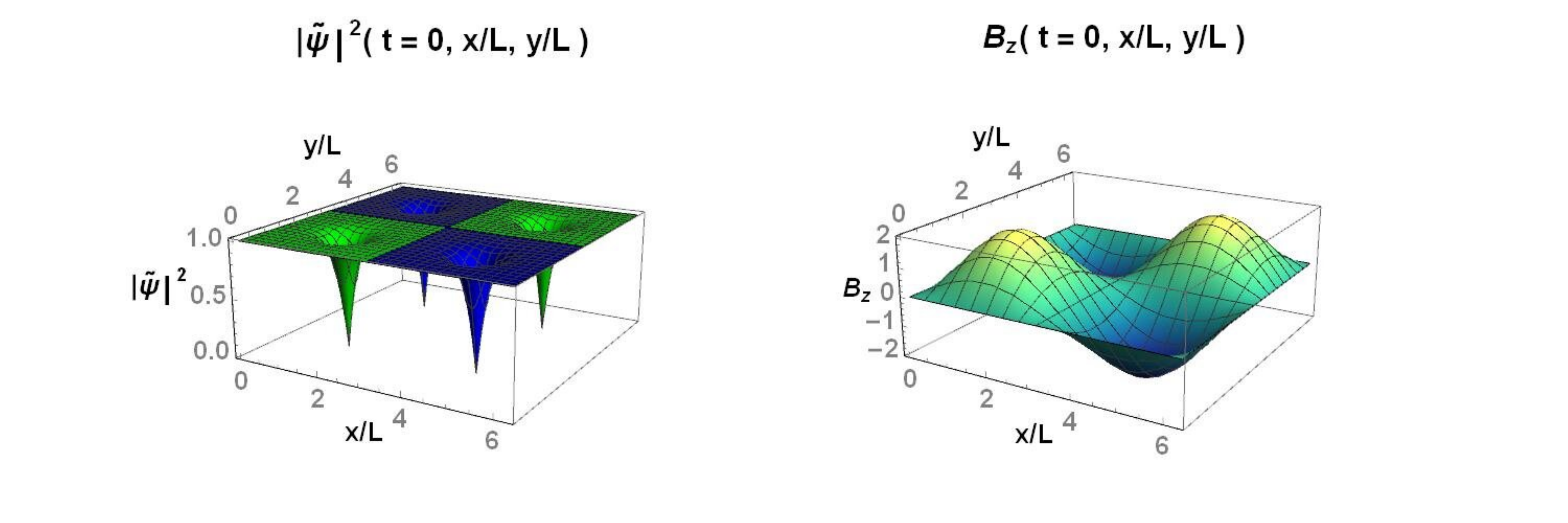}
\caption{\label{fig:plotsteo} \small{Plots of $|\tilde{\psi}(x, y, t=0)|^2$ (left) and magnetic field $B_z(x,y, t = 0)$ (right) on the $2\pi L \times 2\pi L$ simulation box. Notice that since $[\mathbf{a}] = [L]$, $B_z = \nabla\times\mathbf{a}$ is non dimensional. The sign of the vorticity of each vortex matches that of the magnetic field, signalled by the matching colors in each plot}.} 
\end{figure}

Our simulation box has in fact four sub-sectors, $[0,\pi]\times [0,\pi]$, $[\pi,2\pi]\times [0,\pi]$, $[0,\pi]\times [\pi,2\pi]$, and $[\pi,2\pi]\times [\pi,2\pi]$. Note that the field in Eq.~(\ref{eq:initialz}) has one zero in each of these subsectors, associated to the position of a vortex (see Fig.~\ref{fig:plotsteo}). Our ``physical" system is nevertheless only one of these sectors, $[0,\pi]\times [0,\pi]$; the other three can be considered simply as a mathematical trick (or images) to implement periodic boundary conditions in the extended domain. As far as $\lambda_L$, $\xi$, and $l_\eta$ are small, and the positions of the vortices are not too close to the boundary of the domain of interest, our vortices will not feel the effects of the border nor be influenced by the image vortices in the  other subsectors. More vortices can be created initially by taking powers or by applying
the translation operator to Eq.~\ref{eq:initialz}), and multiplying the resulting superconducting order parameters for each individual vortex.

For the vector potential we choose as initial condition:
\begin{eqnarray}
    a_x (x, y, t=0) = a_0 \sin{(x)}\cos{(y)} \label{eq:initialax}, \\
    a_y (x, y, t=0) = a_0 \cos{(x)}\sin{(y)}\label{eq:initialay},
\end{eqnarray}
which sets the initial magnetic field as $B_z (x, y, t=0) = 2 a_0 \sin{(x)}\sin{(y)}$, with $a_0$ a normalization constant related to the magnetic flux (see Eq.~\ref{eq:u0norm}). These initial conditions are known in the area of fluid dynamics as the Taylor-Green vortex, and correspond to the ones used by \cite{nore} as a fixed background in the study of quantized vortices in superfluids.  They do not correspond to any concrete physical realization from the point of view of superconductivity. Nevertheless, as far as we are not concerned with the initial transient, our choice satisfies the correct requirements of periodicity and topology needed for our purposes, and are simple to implement numerically. 

The vector potential is such that the circulation of the magnetic field (related to the winding number or vorticity) in the entire simulation box is 0, but on the $[0,\pi] \times [0,\pi]$ sector it is
\begin{equation}
    \oint B_Z ds = \int_0^\pi \int_0^\pi 2 a_0 \sin{(x)}\sin{(y)}dx\, dy = 8 a_0 .
\end{equation}
It is easy to verify that this calculation yields the same result in the regions $[0,\pi]\times [0,\pi]$ and $[\pi, 2\pi] \times [\pi, 2 \pi]$ and opposite to that in the regions $[0,\pi]\times [\pi,2\pi]$ and $[\pi, 2\pi] \times [0, \pi]$. The sign of the flux in each of the 4 subsectors is correlated to the winding number of the vortex configurations, as is shown in the matching colours of the plots in Fig.~\ref{fig:plotsteo}.

Finally, in order to satisfy the condition that the magnetic flux is conserved throughout the simulation we need to adjust the ratio of the circulation of the magnetic field to the number of vortices. This is done by setting $a_0$ to satisfy that the magnetic flux of $n_v$ vortices in the $[0,\pi] \times [0,\pi]$ sector is $\Phi = n_v \Phi_0 = 4\pi\alpha n_v$, with $\Phi_0$ the flux quantum. Thus, $a_0$ satisfies that:
\begin{equation}
    a_0 = \frac{\pi \alpha n_v}{2\sqrt{2}} .
    \label{eq:u0norm}
\end{equation}

\subsection{Application of the numerical method to the standard Landau Ginzburg problem}
\label{subs:numerical}

In this section we apply our numerical method to the standard GL problem, i.e, with no nematic coupling to the superconductor order parameter. The main purpose of this section is to verify the validity of the numerical scheme by comparing with known results obtained by other methods or from the theory. Readers interested only in the influence of nematicity on the vortex-vortex interaction may skip the following results and refer directly to section \ref{sec:nematicvort}.

From the previous discussion, we remind the reader that the relevant parameter in our theory is the Ginzburg Landau parameter $\kappa$. 
Regarding the dynamics, we have also defined the diffusion constants and the electrical resistivity, which control the rate of the relaxation processes involved in our model. In this work we are not interested in very specific details of the dynamics, but rather in the simpler question of whether equilibrium configurations exist, and in the repulsive or attractive character of the interactions between topological objects. To answer these questions the specific values of the diffusion constants are not relevant, and thus we will choose them all to be of the same order of magnitude.

We will first focus  on the $n_v=1$ vortex problem, taking three representative values for the GL parameter: $\kappa_{-} = 0.49$, $\kappa_c = \frac{1}{\sqrt{2}}$, and $\kappa_{+}=0.92$ (these specific values are chosen to make a comparison with results published before, see below).

\begin{figure}
\centering
\includegraphics[width=1.0\textwidth]{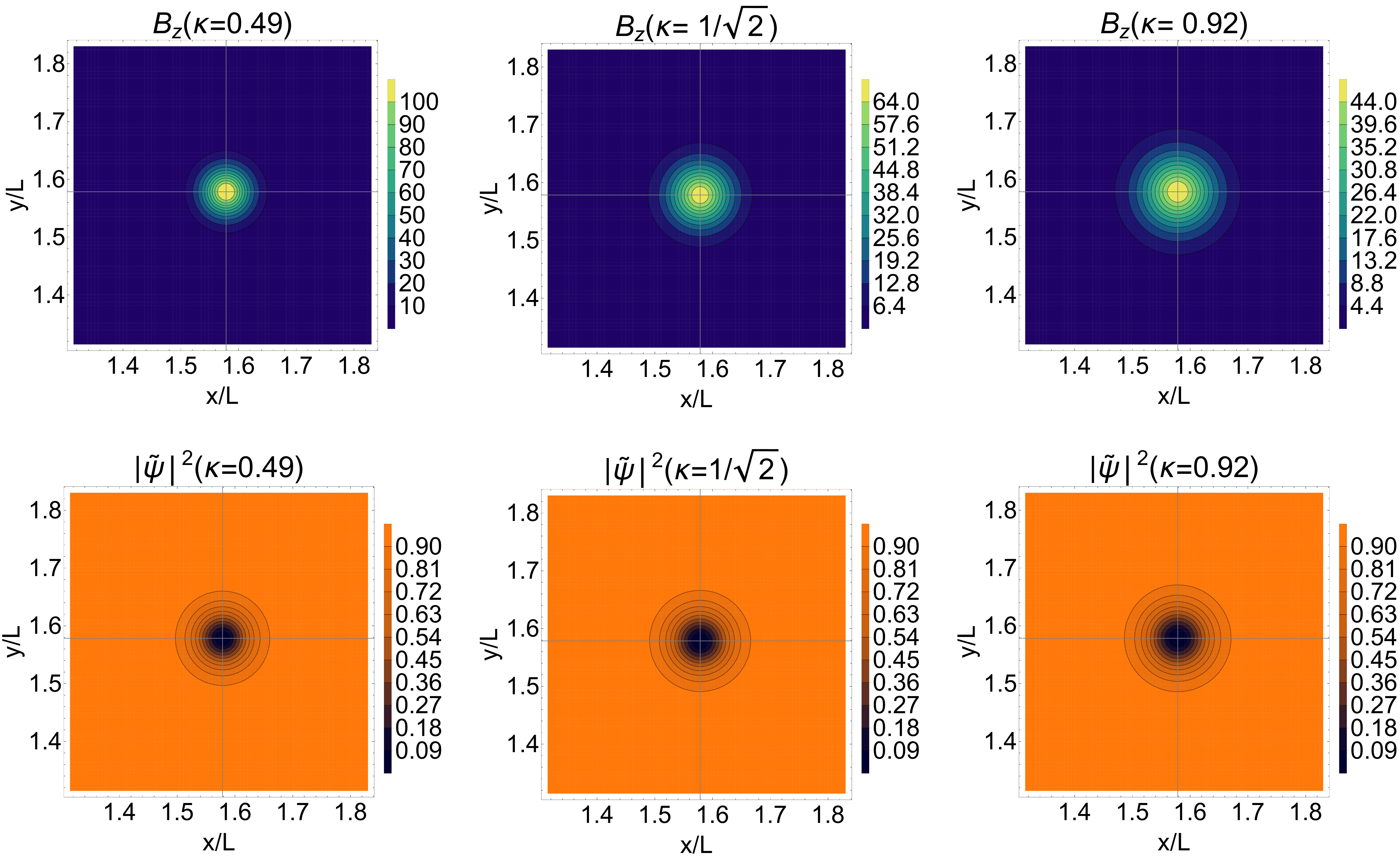}
\caption{\small{Density plots of the magnetic field $B_z$ (top) and of the superconductor order parameter $|\tilde{\psi}|^2$ (bottom). The grey lines serve as visual guides that intersect at $(\pi L/2, \pi L/2)$. A small region in the vicinity of the vortex is shown. The superconductor coherence length is fixed at $\xi = 0.04 L$, which implies that the most noticeable change is in the London length $\lambda_L$. It is smaller than $\xi$ for $\kappa < \kappa_c$ and larger for $\kappa > \kappa_c$, as can be readily seen from the first and third panels.}}
\label{fig:1vGL}
\end{figure}

For the case of of $n_v = 1$ it is known that static cylindrically symmetric solutions exist for any value of $\kappa$. Indeed, it is easy to see that starting from Eq.~(\ref{eq:initialz}) as an initial condition, a static solution is attained after a fast relaxation.
We show in Fig.~\ref{fig:1vGL} density plots of the order parameter and magnetic field for the three chosen values of $\kappa$. Naturally, the equilibrium configurations have cylindrical symmetry and the same results could have been obtained in this particular case by solving a simpler set of ordinary nonlinear differential equations. 

In performing the simulations we maintained a fixed value of $\xi = 0.04 L$ and varied $\kappa$, so the most noticeable effect is on the magnetic field via the change of $\lambda_L$ ($\tilde{\psi}$ also changes, but in a less evident way).

\begin{figure}
    \centering
    \includegraphics[width=1.0\textwidth]{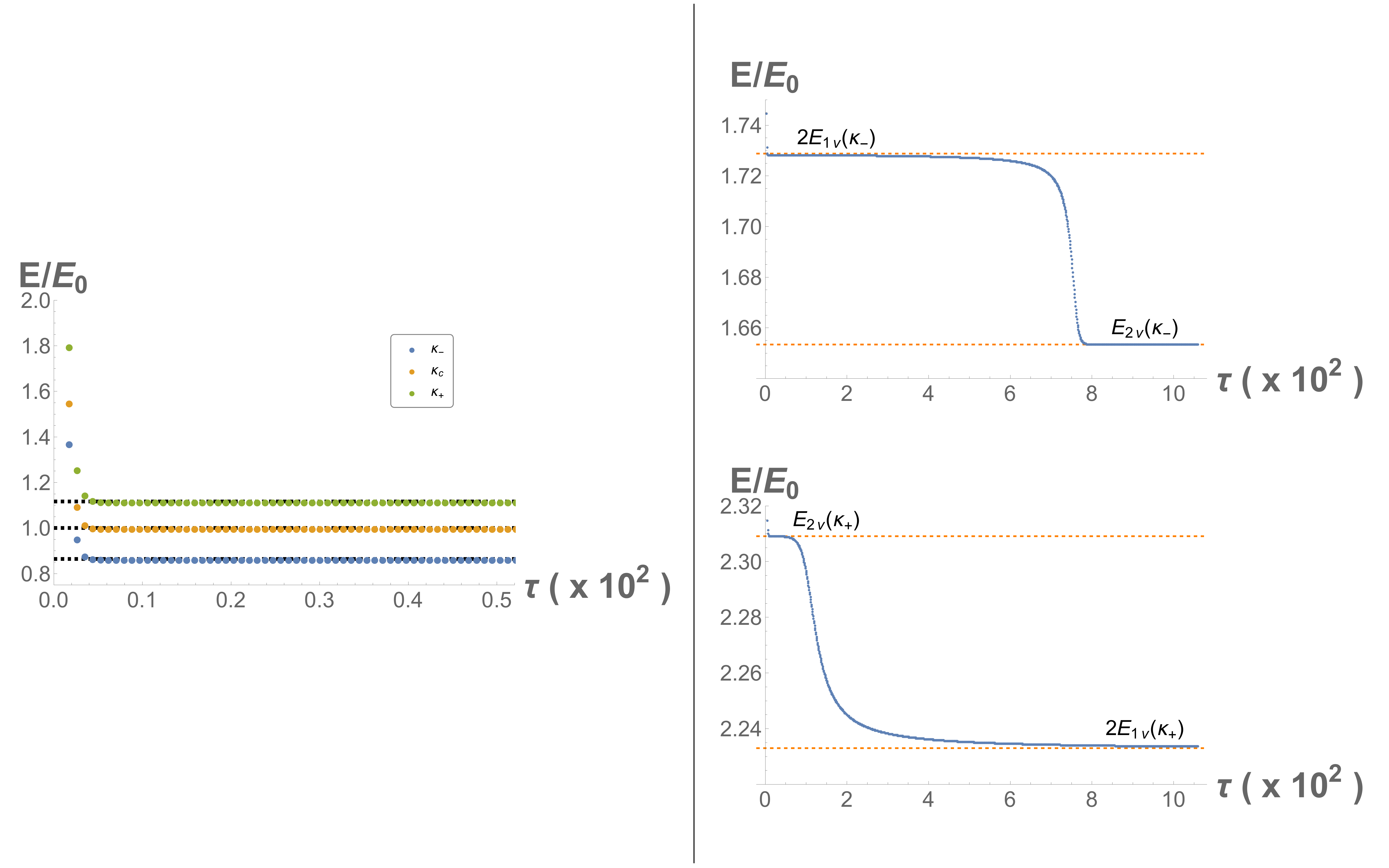}
    \caption{\small{\textit{Left}: normalized and dimnesionless energy, $E/E_0$, as a function of the dimensionless time $\tau$ for a single vortex under TDGL dynamics and 3 different values of $\kappa$. After a fast relaxation, the energy converges to the expected value (dashed lines) according to static numerical calculations using cilindrically symmetric solutions (see Ref.~\cite{jacobs}). \textit{Right}: energy as a function of the physical time $t$ of a two vortex configuration for $\kappa_{-} = 0.49$ (\textit{top}) and $\kappa_{+} = 0.92$ (\textit{bottom}).}}
    \label{fig:energytimev2}
\end{figure}

The fast convergence towards the static configuration is better observed by looking at the time evolution of the energy per unit length of the vortex configuration, as  shown on the left panel of Fig.~\ref{fig:energytimev2} for three chosen values of $\kappa$. It is well known \cite{bogomolny} that at the critical value $\kappa_c$, a configuration of $n_v$ vortices has an energy per unit length  of $E_c = n_v E_0$, with $E_0=\Phi_0/2$ in our notation. We thus use $E_0$ to normalize our results for each $n_v$. For example, a configuration of a single vortex at $\kappa = \kappa_c$ should have a normalized energy of $E/E_0 = 1$, and energy $E < 1$ ($E > 1$) for $\kappa < \kappa_c$ ($\kappa > \kappa_c$).  In all cases, our calculations of the energy converges to a value  which corresponds to the energy of the static configuration (compared with values reported in Ref.~\cite{jacobs}), and is dependent on the value of $\kappa$ as expected. This example shows that the dynamical numerical method can reproduce known results for the single vortex case.


We now turn to study the case of two vortices ($n_v=2$). It is well known that in this case static vortex solutions exist only for $\kappa \leq \kappa_c$, while for $\kappa=\kappa_c$ vortex configurations can exist at arbitrary separations between vortices. For $\kappa<\kappa_c$ a {\em giant} vortex with $n_v=2$ is expected, and for $\kappa>\kappa_c$, due to the repulsive character of vortex-vortex interaction, no static configuration is expected (unless the system is subject to the {\em pressure} of an external magnetic field). We take the initial condition on the superconductor order parameter as
\begin{equation}
    \tilde{\psi}(t=0, x, y) =\tilde{\psi}_v^2 \left[ \frac{(\lambda + i \mu)}{\sqrt{\lambda^2 + \mu^2}}\tanh{\left(\frac{\sqrt{\lambda^2 + \mu^2}}{\sqrt{2}\xi}\right)}\right]^2 ,
\end{equation}
which is just the square of Eq.~(\ref{eq:initialz}). This corresponds to two superimposed single vortices with total vorticity corresponding to $n_v = 2$, referred before to as a {\em giant} vortex. Naturally, the normalization constant $a_0$ of the vector field has to be adjusted for the case $n_v = 2$. 

Note also that the vortices can be initially placed at different positions, respect to each other, by adding or substracting a real constant $d$ to $\lambda$ or $\mu$, which will separate them either in the $x$ o $y$ direction depending on which one is chosen.

\begin{figure}
\centering
\includegraphics[width=1.0\textwidth]{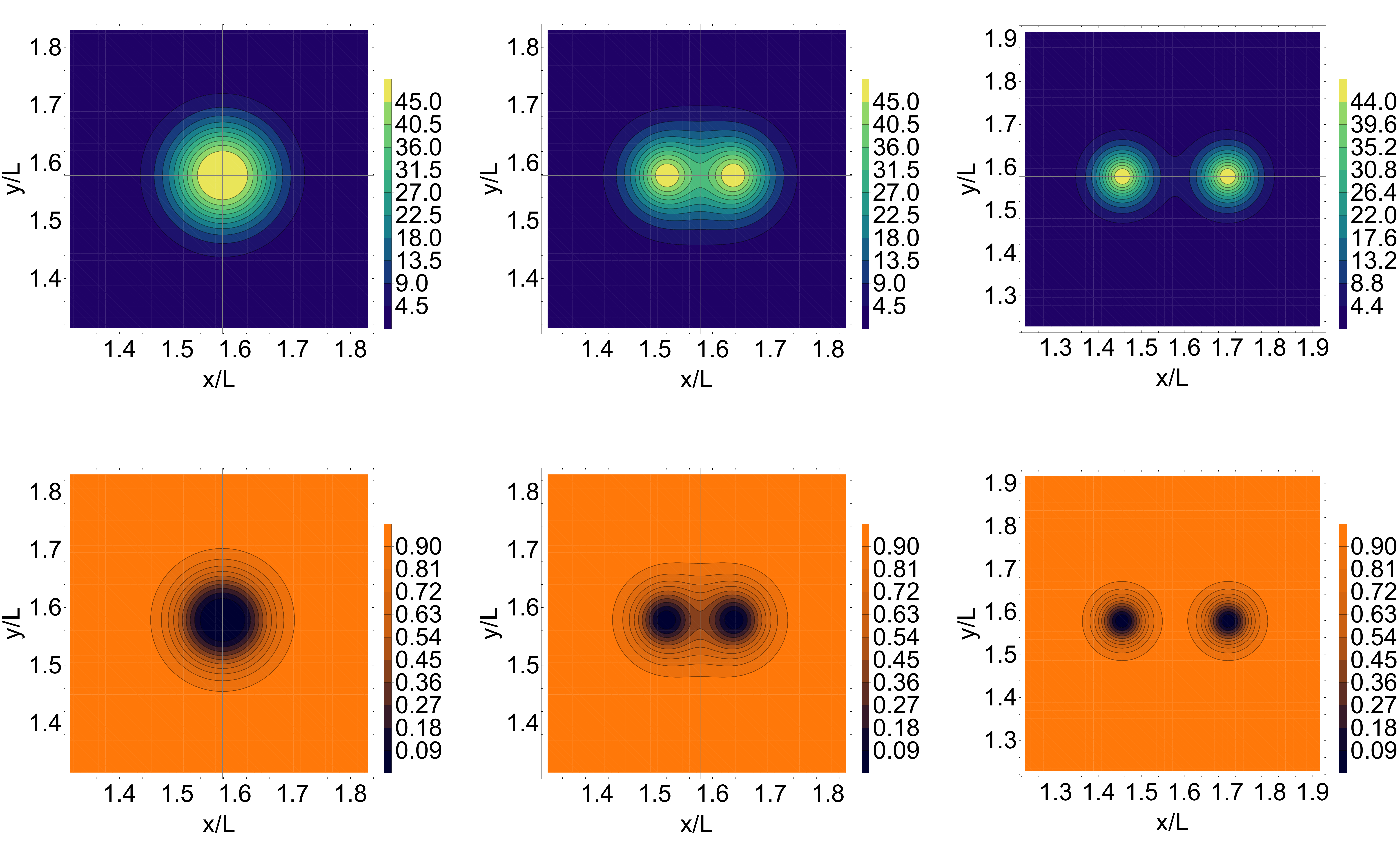}
\caption{\small{Density plots of the magnetic field (top) and of the superconductor order parameter (bottom) for the dynamics of two superimposed $n_v = 1$ vortices (see text) at three different time instances with a GL parameter of $\kappa=0.92$. The superconducting coherence length is fixed at $\xi=0.04 L$.} }
\label{fig:figsplit}
\end{figure}

We show in the right panel of Fig.~\ref{fig:energytimev2} the energy as a function of time for $\kappa = \kappa_{-}$ and $\kappa_{+}$. For $\kappa_{-}$ we placed the vortices at a distance $d = 3.5\xi$ from each other, and let the system evolve. Notice that the total energy of the configuration starts as two times the energy of a single vortex (as reported in Ref.~\cite{jacobs}) for this value of $\kappa$, and as time passes it converges to the energy of two superimposed vortices, as expected. To study $\kappa_{+}$ we set $d = 0$; note that in this case the energy first stays in a plateau and in a second stage starts decreasing again. The first plateau is a transient corresponding to the energy of the \textit{unstable} axially symmetric $n_v = 2$ solution, while the value that the total energy finally converges to corresponds to twice the energy of a single $n_v = 1$ vortex. In Fig.~\ref{fig:figsplit} we show the field configurations at different times in the evolution for $\kappa_{+}$. The two superimposed $n_v = 1$ vortices are not a stable solution of the TDGL equations, and the giant vortex splits into two vortices of vorticity $n_v=1$, also as expected. 

\section{\label{sec:nematicvort} Vortices in the nematic phase}

In this section we will study how the nematic coupling affects the vortex-vortex interaction and their physical properties, focusing first on how coupling of $\tilde{\psi}$ to a nematic order parameter $\tilde{\eta}$ changes what we found in the previous section for the $n_v=1$ single vortex configuration. We will then consider the particular case in which the nematic order parameter is taken as a constant background, showing that self dual equations and the value of the critical GL parameter, separating attracting and repulsing interactions, can be derived exactly by using a simple modification of the original Bogomol'nyi \cite{bogomolny} analysis. Finally, we will relax the hypothesis of a fixed constant nematic background and study how nematicity affects the interactions between vortices in a more general set up. 

\subsection{Single Vortex  with nematic order}

We start by presenting the results for the case in which only a biquadratic coupling between the superconducting order parameter and the nematic order parameter is present, represented by the term proportional to $\hat{\lambda}_2$ in the equations of motion and free energy. The idea is to understand the role of each coupling to the nematic parameter separately, and how each one of them changes the properties of the vortex-vortex interaction.  Notice that in the presence of the biquadratic coupling, both the superconducting coherence length $\xi$ and the nematic coherence length $l_\eta$ (defined when there is no coupling between the order parameters) do not necessarily represent the relevant lengths of the problem. Indeed, ignoring for the moment the magnetic field and under the assumption that a linear approximation is valid, we can write 
\begin{equation}
    \tilde{\psi} = \tilde{\psi_v} + h_1 , \quad \tilde{\eta} = \tilde{\eta_v} + h_2 ,
\end{equation}
where $h_1$ and $h_2$ are small perturbations around the equilibrium values. Substituting into the equations, we get: 
\begin{equation}
    \nabla^2 \begin{pmatrix}
    h_1 \\
    h_2
    \end{pmatrix} = \begin{pmatrix}
    \frac{2\tilde{\psi}_v^2}{\xi^2} & \frac{2\hat{\lambda}_2 \tilde{\psi}_v \tilde{\eta}_v }{\xi^2}\\
    \frac{2\hat{\lambda}_2 \tilde{\eta}_v \tilde{\psi}_v}{\Gamma_2} & \frac{2 \tilde{\eta}_v^2}{l_\eta^2}
    \end{pmatrix} \begin{pmatrix}
    h_1 \\
    h_2
    \end{pmatrix} .
\end{equation}
The eigenvalues $(\alpha_1, \alpha_2)$ of the matrix are related to the solutions for the perturbations, which will be a linear combination of decaying Bessel functions (In the Higgs model version of the model, the inverses of these eigenvalues are related to effective masses for each of the fields).:
\begin{equation}
       h_{1,2} =  C_1 K_0({\sqrt{\alpha_1}(r-r_0)} )+ C_2 K_0({\sqrt{\alpha_2} (r-r_0)} ) .
\end{equation}
For small $\hat{\lambda}_2$ we can do a pertubative expansion, finding
\begin{gather}
   \alpha_1= \frac{2}{l^2} \left(1-\frac{ \hat{\lambda}_2}{\Gamma_4 }-\frac{
   l^2\hat{\lambda}_2^2}{\Gamma_4 (l_\eta^2- \xi ^2)}\right)+O\left(\hat{\lambda}_2^3\right) ,
\\
 \alpha_2=  \frac{2}{\xi ^2}\left(1-\hat{\lambda}_2+\frac{\xi^2 \hat{\lambda}_2^2}{\Gamma_4(
   l_\eta^2- \xi ^2)}\right)+O\left(\hat{\lambda}_2^3\right) ,
\end{gather}
while if $ l_\eta = \xi $
\begin{equation}
\alpha_1= \frac{2 \left(\Gamma_4-\hat{\lambda}_2^2\right)}{\Gamma_4 \xi^2} , \quad \quad
\alpha_2= 
    \frac{2 (\hat{\lambda}_2 - 1) (\hat{\lambda}_2  -\Gamma_4)}{\Gamma_4 \xi^2} .
\end{equation}
We see that for  $\hat{\lambda}_2>0$ ($\hat{\lambda}_2<0)$ the core of the vortex increases (decreases) in size. The  validity of the lineal approximation depends on the value of the London length. Indeed, the linear approximation is not expected to be valid for large $\kappa$, and higher order terms involving the gauge fields need to be retained in order to predict the correct asymptotic behaviour \cite{peri}. For a recent discussion of this issue in a setting similar to ours, see for instance \cite{fideljr}. 

\begin{figure}[ht]
\centering
\includegraphics[width=1.0\textwidth]{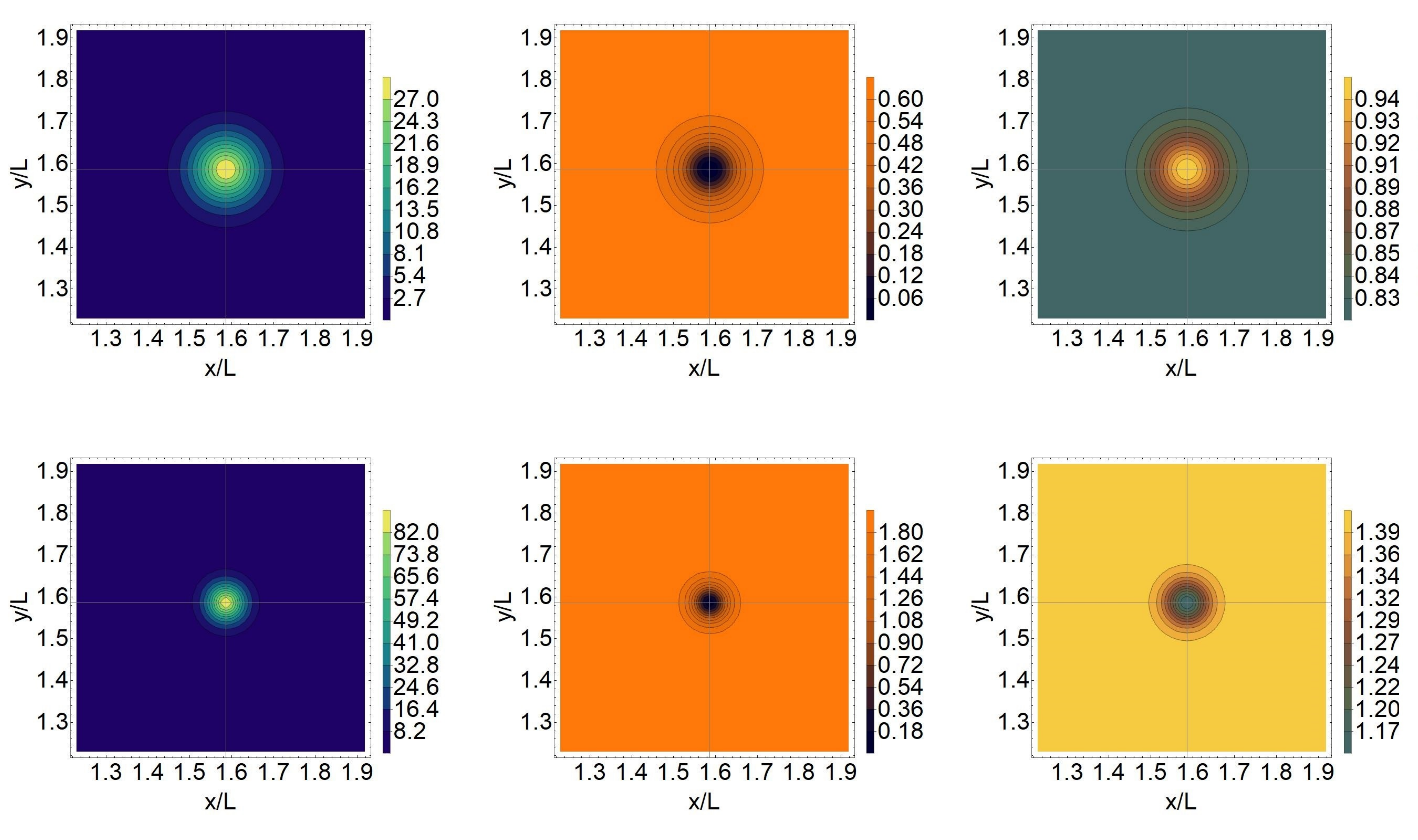}
\caption{\small{Density plots of $B_z$, $|\tilde{\psi}|^2$ , and $\tilde{\eta}$ in the $xy$ plane for $\hat{\lambda}_2 = 0.5$ (\textit{top}) and $\hat{\lambda}_2 = -0.5$ (\textit{bottom}) with $\hat{\lambda}_1 = 0$. The GL parameter in both cases is $\kappa = 0.92$. A small region in the vicinity of the vortex core is shown. Cylindrical symmetry is maintained and the nematic order parameter varies in the vortex core: it is enhanced for $\hat{\lambda}_2 > 0$, and depressed for $\hat{\lambda}_2 < 0$.}}
\label{fig:1v_g1neq_g50}
\end{figure}

The magnetic field and the superconducting and nematic order parameters are displayed in Fig.~\ref{fig:1v_g1neq_g50}, where we show the density plots obtained by solving the full TDGL, using the same initial conditions for $\tilde{\psi}$ and $\mathbf{a}$ as for the standard GL problem (Eqs.~(\ref{eq:initialz}), (\ref{eq:initialax}) and (\ref{eq:initialay})). For the nematic order parameter we chose a uniform background $\tilde{\eta}(t=0,x,y)=\tilde{\eta}_v$.

We notice that for positive coupling, the nematic order parameter is enhanced in the core of the vortex  (and depressed far away from the core), while depressed in the same region for negative $\hat{\lambda}_2$ (and enhanced far away from the core). Also, note that the superconductor order parameters and the magnetic field are more spread out in space for positive coupling than for negative coupling. This is so because the effective coherence length and the effective London length are related to the value of $\tilde{\psi}_v$, which shows this behavior with $\hat{\lambda}_2$. As expected, when $\hat{\lambda}_1 = 0$ the distributions of each order parameter and the magnetic field in the $xy$ plane are cylindrically symmetric. 

\begin{figure}[ht]
\centering
\includegraphics[width=1.0\textwidth]{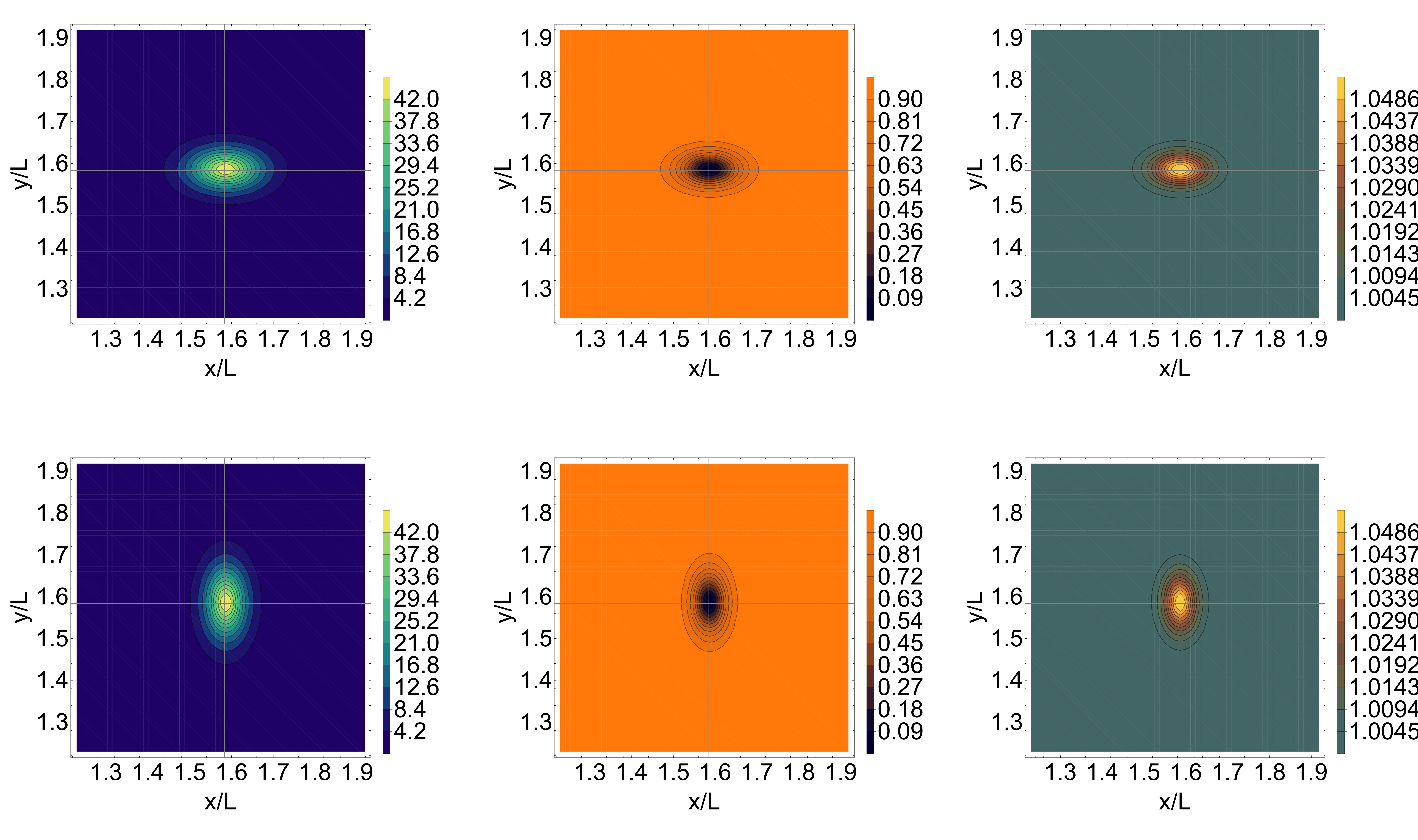}
\caption{\small{Density plots obtained by solving TDGL equations for a single vortex with a $C_4$-symmetry breaking coupling, with $\hat{\lambda}_1 = 0.5$ (\textit{top}) or $\hat{\lambda}_1 = -0.5$ (\textit{bottom}). In both cases the biquadratic coupling $\hat{\lambda}_2$ is set to zero. The vortex cores elongate along a preferred direction, determined by the sign of $\hat{\lambda}_1$.  Notice that for both signs of the coupling parameter the nematic order parameter is enhanced in the vortex core.}}
\label{fig:1v_g10_g5neq}
\end{figure}

As a second step we analyze how the behavior is modified when there is a $C_4$ symmetry breaking coupling, in particular, we consider the two signs with $\hat{\lambda}_1 = \pm 0.5$. As we have stated before, the idea is to better understand the role of each parameter in the vortex-vortex interaction. In particular, $\hat{\lambda}_1$ couples the nematic order parameter to the derivatives of the superconducting order and the magnetic field in a different way, i.e., with a different sign, for the $\hat{x}$ and $\hat{y}$ directions. The resulting density plots from running the TDGL dynamics are shown in Fig.~\ref{fig:1v_g10_g5neq}. The effect of the $C_4$-symmetry breaking coupling causes the vortices to elongate along the $\hat{x}$ ($\hat{y})$  axis for positive (negative) $\hat{\lambda}_1$. Notice nevertheless that in \textit{both} cases the nematic order parameter is enhanced in the vortex core.

\begin{figure}
    \centering
    \includegraphics[width=1.0\textwidth]{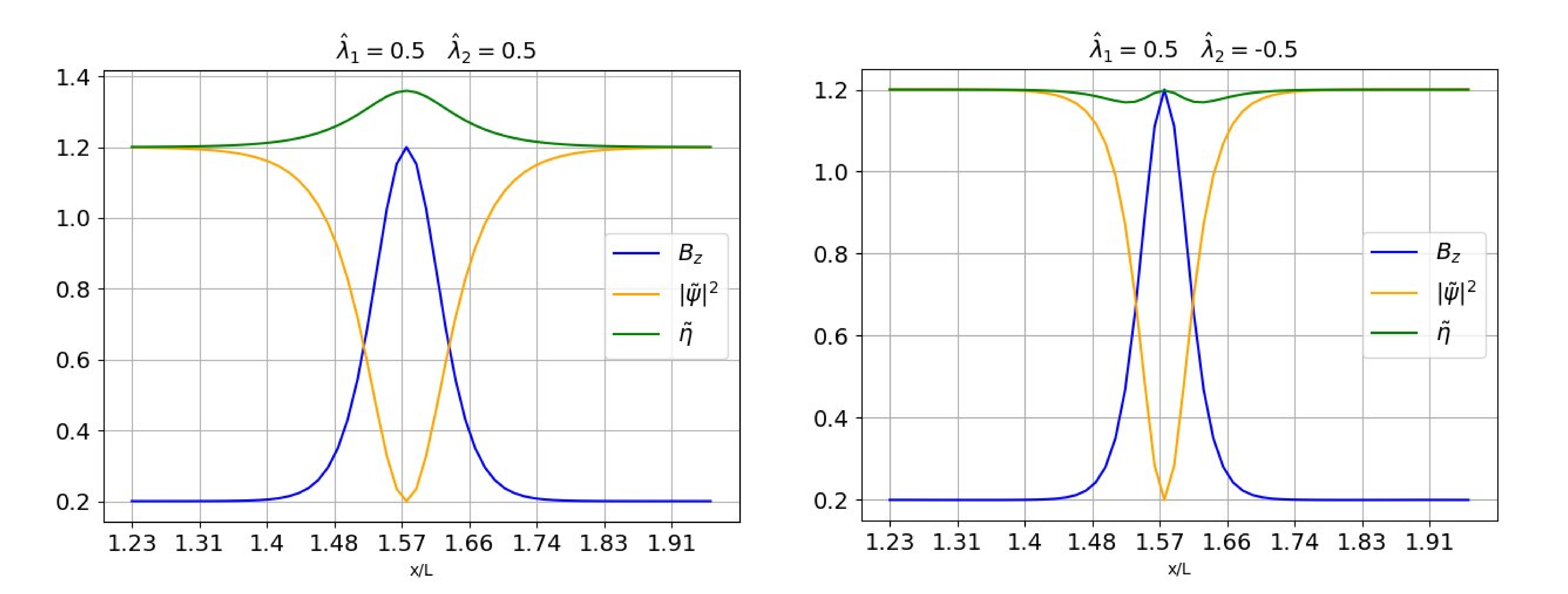}
    \caption{\small{Profiles of the order parameter, magnetic field, and nematic order parameter along $x$ for $y = \pi /2$, with fixed $\lambda_1 = 0.5$ for $\lambda_2 = 0.5$ (left) or $\lambda_2 = -0.5$ (right). The relation between the coherence lengths of the superconductor and nematic order parameter is $l_{\eta}/\xi = 1$. A local maximum of $\tilde{\eta}$ aligns itself with the magnetic field around $x/L = 1.57$ for $\hat{\lambda}_1 = 0.5$ and $\hat{\lambda}_2 = -0.5$ (see right panel).} }
    \label{fig:profiles_g1g5}
\end{figure}

Finally,  we analyze the case when both  $\hat{\lambda}_2 \neq 0$ and $\hat{\lambda}_1 \neq 0$. When $\hat{\lambda}_2 >0$ and $\hat{\lambda}_1 \neq 0$ both couplings tend to enhance the value of the nematic order parameter in the core of the vortex, so we do not expect major surprises. But for the case $\hat{\lambda}_2 < 0$ and $\hat{\lambda}_1 <0$ both terms compete, and the behavior of the nematic order parameter in the vortex core is more difficult to predict. This situation can  be observed in Fig. \ref{fig:profiles_g1g5}. Indeed, for specific values of $\hat{\lambda}_2$ and $\hat{\lambda}_1$  the minimum  of the nematic order parameter may  happen in a ring around the vortex core, as illustrated in Fig.~\ref{fig:profiles_g1g5}. 

\subsection{Self Dual equations with a uniform nematic background}\label{sec:bogomolny}

We show in this section the existence of self dual equations for the case in which the nematic order parameter is taken as a fixed given  {\em constant} background  (in space and time)  $\tilde{\eta}=\tilde{\eta_b}$. Here, the role of the nematic order parameter is to generate a fixed asymmetry in the x-y plane (for a similar phenomenon in multiband superconductors see for instance \cite{babaev_99}). The  proof is a very simple modification of the original reasoning in \cite{bogomolny}, and it helps to understand mainly the influence of $\hat{\lambda}_1$ in $\kappa_c$.
The free energy   can be written as:
\begin{eqnarray}
     F &=& \rho_0 |\alpha_{GL}| \int_V \frac{1}{2}\left(|\tilde{\psi}|^2-1\right)^2 + \xi^2 (1+\hat{\lambda}_1\tilde{\eta}_b) \left|\mathcal{D}_x\tilde{\psi} \right|^2 + \xi^2 (1-\hat{\lambda}_1\tilde{\eta}_b) \left|\mathcal{D}_y\tilde{\psi} \right|^2+  \\ \nonumber && +\frac{\kappa^2}{4}(\nabla \times \mathbf{a})^2 + \hat{\lambda}_2 |\tilde{\psi}|^2 \eta_b^2 .
 \end{eqnarray}
Defining  $\Gamma_x = 1+\hat{\lambda}_1\tilde{\eta}_b$ and $\Gamma_y = 1-\hat{\lambda}_1\tilde{\eta}_b$ we obtain the modified Bogomol'nyi identity,
\begin{gather}
    |\sqrt{\Gamma_x}\mathcal{D}_x \tilde{\psi} -i\sqrt{\Gamma_y}\mathcal{D}_y\tilde{\psi}|^2 = \Gamma_x \mathcal{D}_x\tilde{\psi} \mathcal{D}_x\tilde{\psi}^{*} + \Gamma_y \mathcal{D}_y\tilde{\psi}\mathcal{D}_y\tilde{\psi}^{*}  +i\sqrt{\Gamma_x\Gamma_y}(\mathcal{D}_x\tilde{\psi} \mathcal{D}_y\tilde{\psi}^{*}-\mathcal{D}_x\tilde{\psi}^{*}\mathcal{D}_y\tilde{\psi}) .
\end{gather}
Then, up to a total derivative term, 
\begin{gather}
\xi^2(\Gamma_x|\mathcal{D}_x\tilde{\psi}|^2+\Gamma_y|\mathcal{D}_y\tilde{\psi}|^2) = \xi^2 |\sqrt{\Gamma_x}\mathcal{D}_x \tilde{\psi} -i\sqrt{\Gamma_y}\mathcal{D}_y\tilde{\psi}|^2 -\frac{\sqrt{\Gamma_x\Gamma_y}}{2}|\tilde{\psi}|^2\nabla\times\mathbf{a} .
\end{gather}
On the other hand, 
\begin{gather}
    \frac{\kappa^2}{4}(\nabla\times\mathbf{a})^2=\frac{\kappa^2}{4}[(\nabla\times\mathbf{a}-c_1(|\tilde{\psi}|^2-c_2))^2  +2c_1(|\tilde{\psi}|^2-c_2)\nabla\times\mathbf{a}  - c_1^2(|\tilde{\psi}|^2-c_2)^2] .
\end{gather}
Then, 
\begin{align}
   F &= |\alpha_{GL}|\rho_0 \int_V \left [  \xi^2 |\sqrt{\Gamma_x}\mathcal{D}_x \tilde{\psi} -i\sqrt{\Gamma_y}\mathcal{D}_y\tilde{\psi}|^2  + \frac{\kappa^2}{4} (\nabla\times\mathbf{a} - c_1 (|\tilde{\psi}|^2 - c_2)) ^2  -\frac{c_1c_2\kappa^2}{2}(\nabla\times\mathbf{a}) \right.\nonumber \\
  & \left.
  + \frac{1}{2}(c_1\kappa^2-\sqrt{\Gamma_x\Gamma_y}) 
  |\tilde{\psi}|^2(\nabla\times\mathbf{a})
 + \frac{1}{2}\left(|\tilde{\psi}|^2-1\right)^2 -\frac{\kappa^2 c_1^2}{4}(\tilde{\psi}^2-c_2)^2 + \hat{\lambda}_2 |\tilde{\psi}|^2  \tilde{\eta}_b^2\right] .
\end{align}
Then, choosing 
\begin{gather}
c_1\kappa_c^2 = \sqrt{\Gamma_x \Gamma_y} , \quad\quad
    \frac{1}{2} = \frac{\kappa_c^2  c_1^2}{4} , \quad\quad
    \frac{\kappa_c^2 c_1^2 c_2}{2} = 1 -\hat{\lambda}_2 \tilde{\eta}_b^2 ,
    \label{eq:bogo2}
\end{gather}
the free energy can be expressed as a sum of squares plus a term proportional to the magnetic flux. Thus, the minimum  energy configurations are found by demanding the squares to be zero, that is
\begin{align}
    &\sqrt{\Gamma_x}\mathcal{D}_x \tilde{\psi} = i\sqrt{\Gamma_y}\mathcal{D}_y\tilde{\psi}, \\
    &\nabla\times\mathbf{a} = c_1(\tilde{\psi}^2-c_2) .
\end{align}
Solving the systems defined in Eq.~(\ref{eq:bogo2}) yields the values of the constants
\begin{equation}
    c_1 = \frac{2}{\sqrt{\Gamma_x \Gamma_y}} , \quad\quad c_2 = 1 - \hat{\lambda}_2\tilde{\eta}_b^2 ,
\end{equation}
and the critical value of the GL parameter
\begin{equation}
    \quad
    \kappa_c^2 = \frac{\Gamma_x \Gamma_y}{2}=\frac{(1-\hat{\lambda}_1^2 \tilde{\eta}_b^2)}{2} .
    \label{eq:bogocrit}
\end{equation}

It is clear that if $\hat{\lambda}_1 = 0$ we recover the classical GL limit where the inter vortex interaction changes character. Thus we see that the main role of $\hat{\lambda}_1$ is to lower $\kappa_c$ with respect to the standard GL theory. Notice that at this step, $\hat{\lambda}_2$ does not affects the value of $\kappa_c$. As we will see in next section, this situation changes once the nematic order parameter becomes dynamical. 

Finally, using the obtained parameters we can calculate that the energy per unit length at $\kappa_c$ is
\begin{equation}
    F = \frac{\Phi_0}{2}(1 - \hat{\lambda}_2\tilde{\eta}_b^2) \sqrt{1-\hat{\lambda}_1^2\tilde{\eta}_b^2} .
\end{equation}
As expected the free energy, is proportional to the magnetic flux (indicating the absence of interaction between vortices) and reduces to the standard Bogomol'nyi result for $\hat{\lambda}_{1,2}=0$.

\subsection{Numerical study of the vortex-vortex interaction}

As we have seen, when there is no coupling to a nematic parameter, it can be deduced from GL theory that $\kappa_{c} = 1/\sqrt{2}$ is the critical value which determines whether the interaction between vortices is attractive or repulsive and, if the nematic order parameter is a constant background, the $C_4$ symmetry breaking coupling $\hat{\lambda}_2$ lowers this value. In this section we solve the full TDGL equations in order to study how $\kappa_c$ changes when we include a coupling between the superconducting and nematic order parameters.
In order to do so, we will start with two vortices placed in the $xy$ plane at a given distance $d$ (of order $\xi$), and study how they evolve under TDGL dynamics. Choosing an initial condition where vortices are already separated is more efficient, as we do not have to wait for the splitting time to see if the configuration is stable or not, a process which can take an extremely long time near $\kappa_c$.

The initial conditions for the nematic order parameter and the vector potential remain the same as in the previous section, but with the condition that the total magnetic flux corresponds to that of two flux quanta. The superconducting order parameter is initially set as:
\begin{gather}
    \tilde{\psi}(t = 0, x, y) = \tilde{\psi}_1 (t=0, x, y) \tilde{\psi}_2 (t = 0, x, y),  \\
    \tilde{\psi}_{1,2} (t=0, x, y) = \tilde{\psi}_v \frac{((\lambda \pm  d) + i \mu)}{\sqrt{(\lambda \pm d)^2 + \mu^2}}\tanh{\left(\frac{\sqrt{(\lambda\pm d)^2 + \mu^2}}{\sqrt{2}\xi}\right)} ,
\end{gather}
which places two $n_v=1$ vortices in the $xy$ plane separated a distance $\approx 2d$ from each other in the $x$ axis. The initial direction in which they are separated can be easily changed to the $y$ axis by choosing to shift $\mu$ instead of $\lambda$. Under the presence of a nematic order, these cases do not need to be equivalent. 

\subsubsection{Biquadratic coupling ($\hat{\lambda}_2 \neq 0, \hat{\lambda}_1 = 0$)}
    
We begin by studying the case for a biquadratic coupling between the superconducting and nematic order parameters. As we have mentioned before, this coupling does not unveil the specific nematic nature of the order parameter, as a biquadratic coupling of this kind could be present for any standard real scalar (i.e, invariant) order parameter too. 
From Eq.~(\ref{eq:range_gam1}) we know that the allowed values of $\hat{\lambda}_2$ are limited by the nematic potential coefficient, $\Gamma_4$. In our simulations we have explored a few representative values of $\Gamma_4$ which in turn define the interval in which $\hat{\lambda}_2$ can vary.

\begin{figure}
\centering
\includegraphics[width=1.0\textwidth]{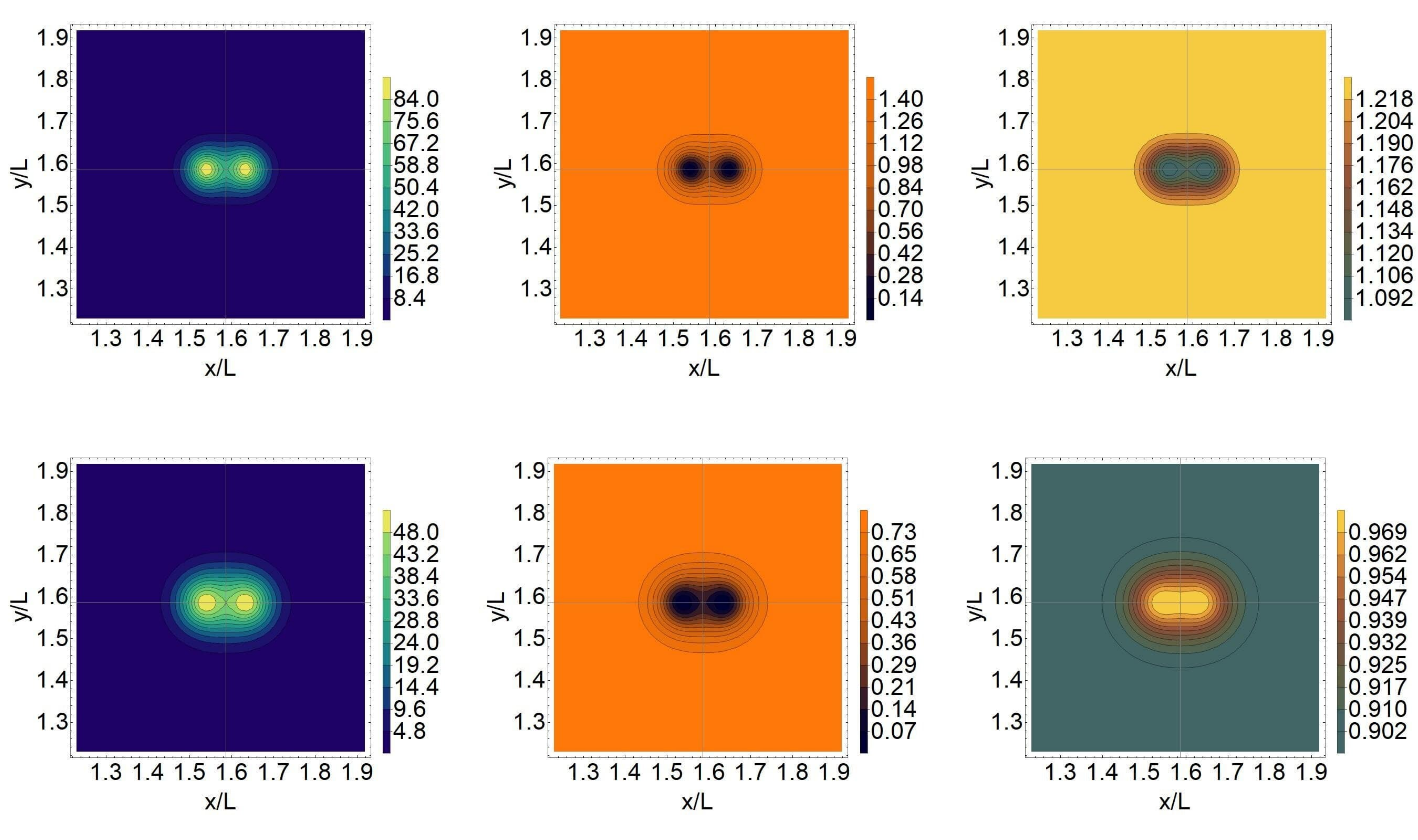}
    \caption{\small{Density plots of the magnetic field, superconducting order parameter, and nematic order parameter for $\hat{\lambda}_2 = -0.35$ (top) and  $\hat{\lambda}_2 = 0.240$ (bottom). Notice that for $\hat{\lambda}_2 < 0$, $\eta < \eta_v$ on the core of the vortices, and $\eta > \eta_v$ for $\hat{\lambda}_2 > 0$.} }
\label{fig: profile_dens1}
\end{figure}

We present first some examples of our simulations starting with two vortices in the $xy$ plane, picking up two opposite sign values for the biquadratic coupling. In Fig.~\ref{fig: profile_dens1} we set $\Gamma_4 = 0.5$ and $\frac{l_\eta^2}{\xi^2} = 0.5$. We fixed the coherence length at $\xi = 0.04 L$ and we varied $\lambda_L$. For each simulation we evaluated whether the vortices attracted or repelled each other by plotting the density of each order in the $xy$ plane and tracking the vortex cores in time. By fine tuning $\lambda_L$ we can determine $\kappa_c$ within a certain margin of error. 

In Fig.~\ref{fig: profile_dens1} (top) we show the density plots for $\hat{\lambda}_2 = -0.35$ and $\kappa_c \cong 0.85$. We chose this as the critical value of the GL parameter, determined by noting that after a long simulation ($\tau=10 \times 10^2$) and for $\kappa = 0.84$ the vortices show very little attraction, while for $\kappa = 0.86$ they show very little repulsion. Therefore, we can estimate that the critical value is $\kappa_c = 0.85 \pm 0.01$. We also verified that for a long simulation there was no resolvable motion of the vortices. The fact that $\kappa_c > 1/\sqrt{2}$ indicates that the biquadratic coupling is inducing an attractive interaction as the region of Type I superconductivity is enlarged. 

In the bottom half of Fig.~\ref{fig: profile_dens1} we show a similar situation but for $\hat{\lambda}_2 = 0.24$ and $\kappa_c \cong 0.74$. The determination of the criticality of this value was determined as before. As with the previous case, the value $\kappa=\kappa_c >1/\sqrt{2}$ is larger than in a standard superconductor. Then, the biquadratic coupling induces an attractive interaction regardless of its sign. 

We explored how these results are affected by the variation of the other parameters and we show some results in table \ref{tb:kc_biquad}. For the range we have explored, $\kappa_c$ does not show a strong dependence on $l_\eta$. Nevertheless, we can see that for a fixed $\hat{\lambda}_2$, $\kappa_c$ approaches $1/\sqrt{2} $ as $\Gamma_4$ increases. This seems reasonable, since as $\Gamma_4$ becomes larger the back reaction of superconductivity on nematicity becomes negligible. Thus, we can expect a fixed and constant nematic order to become a better approximation, also as we have already seen via the self dual equations that $\kappa_c$ does not depend on $\hat{\lambda}_2$ in this limit.
\subsubsection{$C_4$ symmetry breaking coupling ($\hat{\lambda}_2 = 0$, $\hat{\lambda}_1 \neq 0$)} 
As mentioned before, the terms proportional to $\hat{\lambda}_1$ in the free energy act by breaking the symmetry between the $x$ and $y$ directions. As we have already discussed, one of the main effects of $\hat{\lambda}_1$ on the fields is to elongate the vortices in a direction that depends on the sign of the coupling parameter. In our simulations, for $\hat{\lambda}_1 > 0$ the vortices elongate in the $x$ direction while for $\hat{\lambda}_1 < 0$ they elongate in the $y$ direction, as has been shown in Fig.~\ref{fig:1v_g10_g5neq}. This coupling, contrary to what happened with $\hat{\lambda}_2$, has the effect of enhancing  the value of the nematic order parameter in the vortex cores regardless of the sign of $\hat{\lambda}_1$.

\begin{figure}
    \centering
    \includegraphics[width=0.9\textwidth]{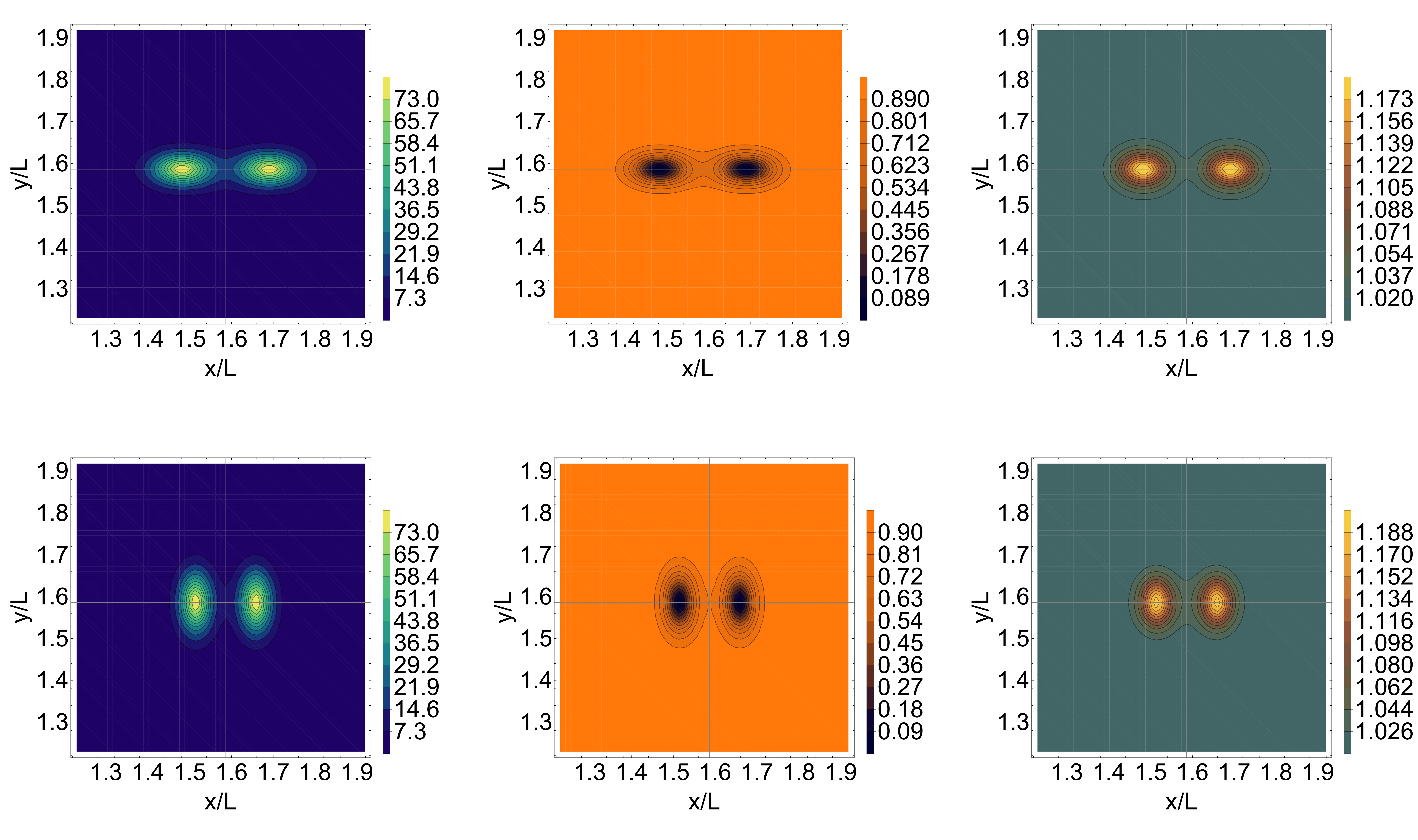}
    \caption{\small{Density plots of the magnetic field ({\em left}), superconducting order parameter ({\em center}), and nematic order parameter ({\em right}) for $\hat{\lambda}_1 = 0.5$ ({\em top} panels) and $\hat{\lambda}_1 =- 0.5$ ({\em bottom} panels). For $\hat{\lambda}_2 = 0$ the equilibrium value for the nematic order parameter is $\tilde{\eta_v} = 1$, and a local maximum for $\eta$ can be found where the vortex cores are, regardless of the sign of the coupling parameter. Also, we can see how the coupling has the effect of elongating the vortices along a selected direction, as was also shown previously in the single vortex case.}}
    \label{fig:dens_gam5}
\end{figure}

We have seen in section \ref{sec:bogomolny} that, in the case of a constant nematic order parameter, the effect of $\hat{\lambda}_1$ is to \textit{decrease} the value of $\kappa_c$, meaning that it mediates a repulsive interaction.   
For a dynamical nematic order parameter, the behaviour of $\kappa_c$ needs to be investigated numerically. Remember that the uncoupled nematic coherence length is defined as $l_\eta^2 = \Gamma_2/\Gamma_3$ and that the superconductor coherence length is fixed at $\xi = 0.04 L$. Therefore, we chose values of $\Gamma_2$ and $\Gamma_3$ that assert that the ratio $l_\eta^2 / \xi^2$ remains of order 1. Within the (high order) accuracy of our numerical method we do not observe a strong dependence either with the sign of $\hat{\lambda}_1$, the value of the nematic coherence length, or $\Gamma_4$, involved in the nematic potential. As hinted by the self dual case, $\kappa_c$ is lower than $1/\sqrt{2}$ and for $\hat{\lambda}_1 = \pm 0.5$ the self dual point gives $\kappa_c = 0.61$. This prediction matches the results obtained by full TDGL dynamics indicating that, within our numerical precision, there is no apparent dependence on any of the aforementioned parameters. The density plots for each relevant variable are shown in Fig. \ref{fig:dens_gam5}.

\begin{table}
\setlength{\tabcolsep}{5pt}
\centering
\begin{tabular}{|c|c|c|c||}
\hline
\multicolumn{4}{|c|}{$\hat{\lambda}_2 = -0.353$} \\
\hline
$\Gamma_4$ & $\frac{l_\eta^2}{\xi^2}=0.5$ & $\frac{l_\eta^2}{\xi^2}=1.0$ & $\frac{l_\eta^2}{\xi^2}=2.0$  \\
\hline
0.25 & 1.05 $\pm$ 0.05 & 1.05 $\pm$ 0.05 & 1.05 $\pm$ 0.05 \\
\hline
0.5 & 0.85 $\pm$ 0.01 & 0.87 $\pm$ 0.01 & 0.88 $\pm$ 0.01 \\
\hline
1 & 0.76 $\pm$ 0.01  & 0.77 $\pm$ 0.01 & 0.78 $\pm$ 0.01 \\ 
\hline
2 &  0.74 $\pm$ 0.01 & 0.74 $\pm$ 0.01  & 0.74 $\pm$ 0.01 \\
\hline
\end{tabular}
\hspace{1mm}
\begin{tabular}{|c|c|c|c||}
\hline
\multicolumn{4}{|c|}{$\hat{\lambda}_2 = 0.240$} \\
\hline
$\Gamma_4$ & $\frac{l_\eta^2}{\xi^2}=0.5$ & $\frac{l_\eta^2}{\xi^2}=1.0$ & $\frac{l_\eta^2}{\xi^2}=2.0$  \\
\hline
0.25 & 0.81 $\pm$ 0.01 & 0.81 $\pm$ 0.01 & 0.80 $\pm$ 0.01 \\
\hline
0.5 & 0.74 $\pm$ 0.01 & 0.76 $\pm$ 0.01 & 0.77 $\pm$ 0.01 \\
\hline
1 & 0.73 $\pm$ 0.01  & 0.73 $\pm$ 0.01 & 0.73 $\pm$ 0.01 \\ 
\hline
2 &  0.72 $\pm$ 0.01 & 0.72 $\pm$ 0.01  & 0.72 $\pm$ 0.01 \\
\hline
\end{tabular}
\caption{\small{Estimated values of the critical GL parameter $\kappa$ when only a biquadratic coupling is present, shown for two representative values of opposite sign of $\hat{\lambda}_2$. There is no clear dependence on the ratio of the coherence lengths. The critical value decreases as we increase $\Gamma_4$}.}
\label{tb:kc_biquad}
\end{table}

\subsubsection{Combined couplings ($\hat{\lambda}_2 \neq 0$, $\hat{\lambda}_1 \neq 0$)}
    
\begin{figure}
\centering
\includegraphics[width=0.7\textwidth]{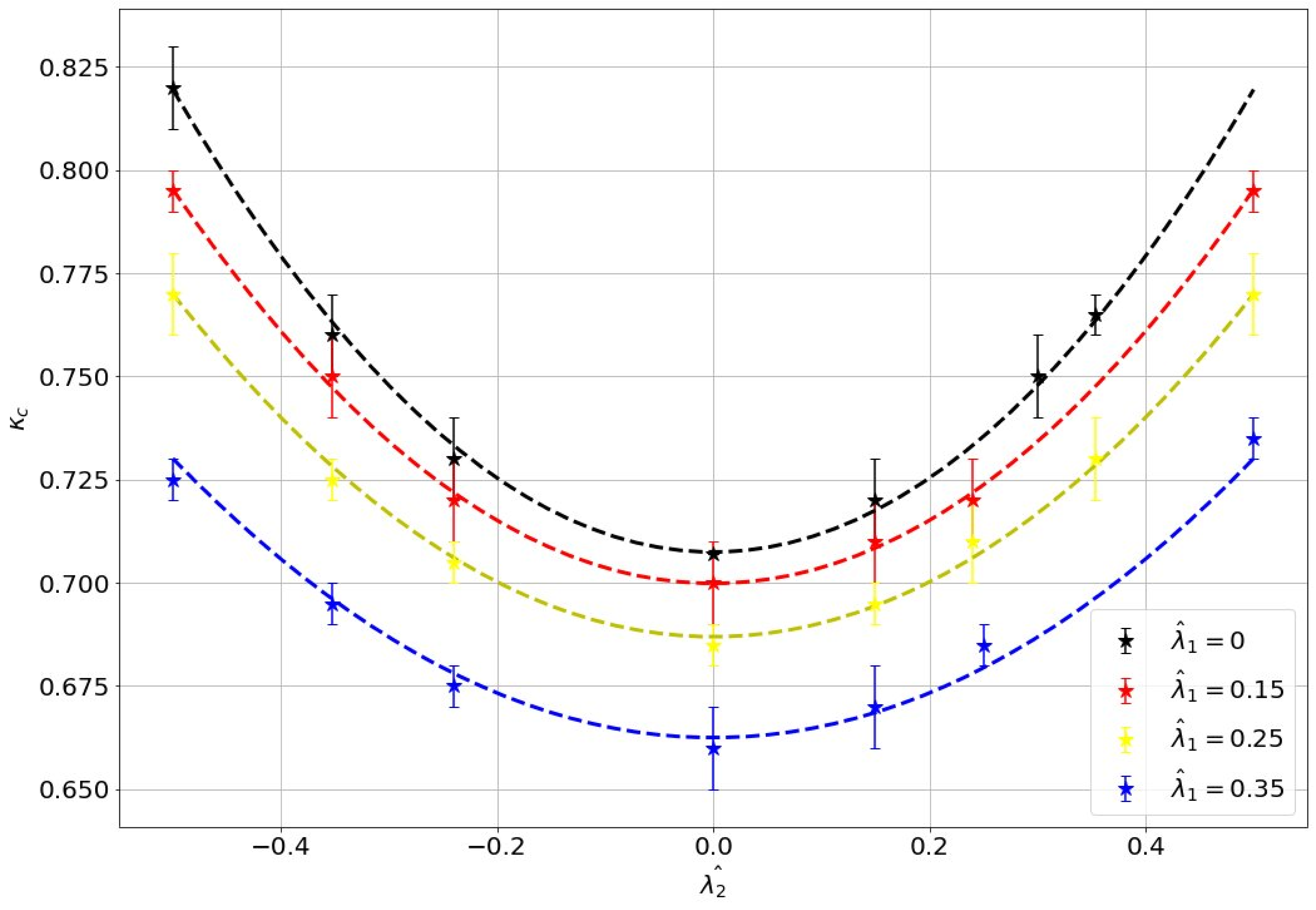}
\caption{\small{Dependence on $\hat{\lambda}_2$ of the critical value of the GL parameter $\kappa_c$ for different values of $\hat{\lambda}_1$. The (blue, orange, green, red) dots correspond to $\hat{\lambda}_1 = (0,0.15,0.25,0.35)$. A simple quadratic model is fitted for each curve, with residuals of order $10^{-3}$. The coefficient multiplying $\hat{\lambda}_2^2$ is decreasing as a function of $\hat{\lambda}_1$.}}
\label{fig:both_couplings}
\end{figure}

We ran simulations turning on both coupling parameters and studied how the critical value of the GL parameter changes in these scenarios. In particular, we fixed the value of $\Gamma_4 = 1$ which in turn defines the allowed values of the biquadratic coupling parameter. Since the dependence on the ratio of the coherence lengths in the previous cases was practically negligible, we fixed $\Gamma_2$ so that the ratio between the uncoupled coherence lengths is 0.5. Having made these choices, we ran simulations for different values of $\hat{\lambda}_1$ and studied the value of the critical GL parameter as a function of $\hat{\lambda}_2$. The value of $\kappa_c$ as a function of $\hat{\lambda}_2$ is presented in Fig.~\ref{fig:both_couplings} for different choices of $\hat{\lambda}_1$.
Within the range of parameters studied, we can observe that the dependence of $\kappa_c$ is quadratic on $\hat{\lambda}_2$, and that the main effect of $\hat{\lambda}_1$ is to lower the value of $\kappa_c$. We can fit the curves with a simple quadratic model
\begin{equation}
   \kappa_c = \kappa_{c,0}(\hat{\lambda}_1) + a(\hat{\lambda}_1)\hat{\lambda}_2^2 .
\end{equation}
For the data shown in Fig~\ref{fig:both_couplings}, $a(\hat{\lambda}_1) \cong 0.45 - 0.5 \hat{\lambda}_1 $, and
\begin{equation}
    \kappa_{c,0}(\hat{\lambda}_1) \cong \frac{1}{\sqrt{2}}(1-\frac{1}{2}\hat{\lambda}_1^2 \tilde{\eta}_v^2) ,
\end{equation}
which is consistent with Eq.~(\ref{eq:bogocrit}). We thus see clearly from Fig.~\ref{fig:both_couplings} the role the two couplings play in the vortex-vortex interaction: while the bi-quadratic coupling induces an attractive interaction stabilizing a type I phase, the term proportional to $\hat{\lambda}_1$ induces a repulsive interaction favouring the formation of a type II phase.
\section{Discussion and Conclusions }
In this work we have analized some of the consequences of a nematic coupling  on the superconductor vortex structure as well as on  the nature of vortex-vortex interactions,  in the framework of Ginzuburg Landau theories,  where nematicity is taken into account by introducing a real order parameter which couples to the complex order parameter (and to the magnetic vector potential) via two terms. The biquadratic term, as expected, may introduce a competitive or a cooperative coupling; we have shown that, in any case,  this term  induces an attractive vortex-vortex interaction and then conspires against the existence of the mixed phase. On the other hand,  the trilinear term, which couples the nematic order parameter to the (covariant) derivatives of the complex superconducting parameter,  always  induces a repulsive interaction that favours the stability of the mixed phase.

A distinctive feature of our work concerns the method used to study the problem of the vortex-vortex interaction. Unlike previous works that use different approximation schemes, as perturbative calculations or variational methods, we have tackled the problem by using a dynamical method based on the solution of the TDGL equations with very high order approximations that converge exponentially fast to the solutions. In particular, we used a numerical technique based on spectral methods. This class of methods, well known in the area of fluid dynamics and other areas of research when solving partial differential equations, are less popular in the study of superconducting materials. 
It is also worth remarking that the method introduces no numerical dispersion or dissipation, thus allowing for precise determination of, e.g., critical values from numerical simulations. In particular, it allowed estimation of the critical value of the Ginzburg Landau parameter $\kappa_c$,  in the case of a superconductor  with combined nematic couplings $\hat{\lambda}_2 \neq 0$ and $\hat{\lambda}_1 \neq 0$, a problem that  can only be studied numerically. We show that the numerical solution recovers the analytical case in the limit of a hard nematic parameter,  where $\kappa_c$ can be approximated with a simple quadratic model on the coupling coefficients.

Because these methods are more stable and much more resource efficient than  finite differences method, which are often applied in the area, it is tempting to explore  their applicability  far beyond the particular problem we have addressed in this work.
Having established the bases of the method, we can envisage many different problems that could be studied using the same techniques. 

An interesting problem that we have already started to consider is the interaction of vortices with nematic domain walls or twin boundaries. As we have already mentioned,  it is natural to expect that in real situations the sample will have twin boundaries, and characterizing this interaction is obviously an interesting question.

\begin{figure}[ht]
    \centering
    \includegraphics[width=0.9\textwidth]{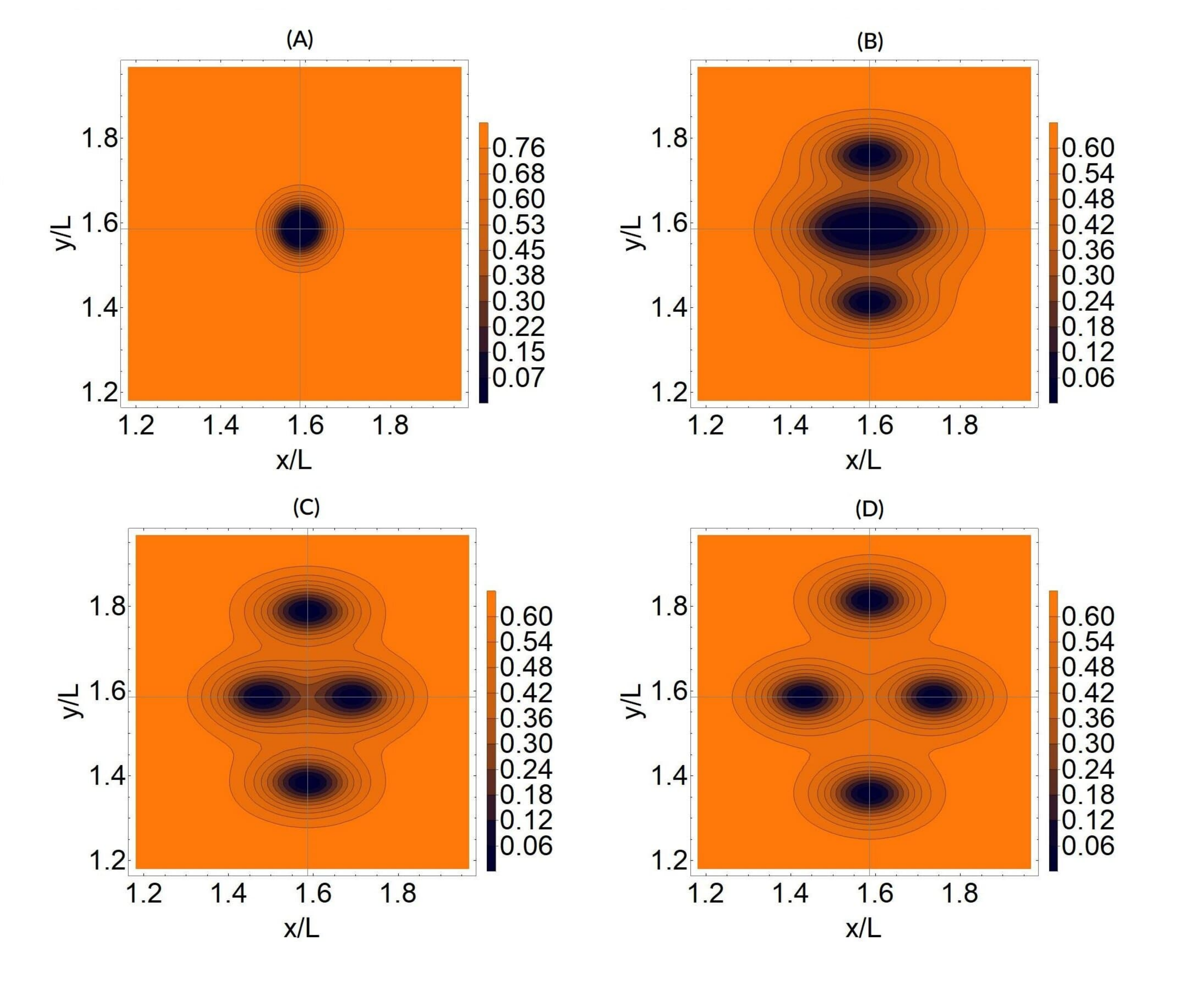}
    \caption{\small{Density plot of the order parameter $|\tilde{\psi}|^2$ for a configuration starting with 4 superimposed $n_v = 1$ vortices with couplings $\hat{\lambda}_1 = 0.5 = \hat{\lambda}_2$} at different times in the evolution (time increasing from panels (A) to (D)). The GL parameter is $\kappa = 1.5$.}
    \label{fig:supernem}
\end{figure}
Moreover, the methods discussed here can be trivially extended to the case of many vortices. Indeed, only by adjusting the vorticity of the initial configuration, the code is able to give us the dynamics of an arbitrary number of vortices. As an example, we show in Fig.~\ref{fig:supernem} the time evolution of a configuration starting with 4 superimposed $n_v = 1$ vortices with both couplings to the nematic order parameter. Large $n_v$ configurations will be useful in the study of Abrikosov lattice formations \cite{obl_exp,oblicuos}.  Once appropriate boundary conditions are implemented \cite{confinantes} the extension to larger values of $\kappa$, typical in Fe-based superconductors, can be considered. As one of the couplings induces an attractive interaction while the other a repulsive one, it is not evident a priori which is the resulting combined effect in the structure of the lattice. 

\begin{figure}
    \centering
    \includegraphics[width=0.65\textwidth]{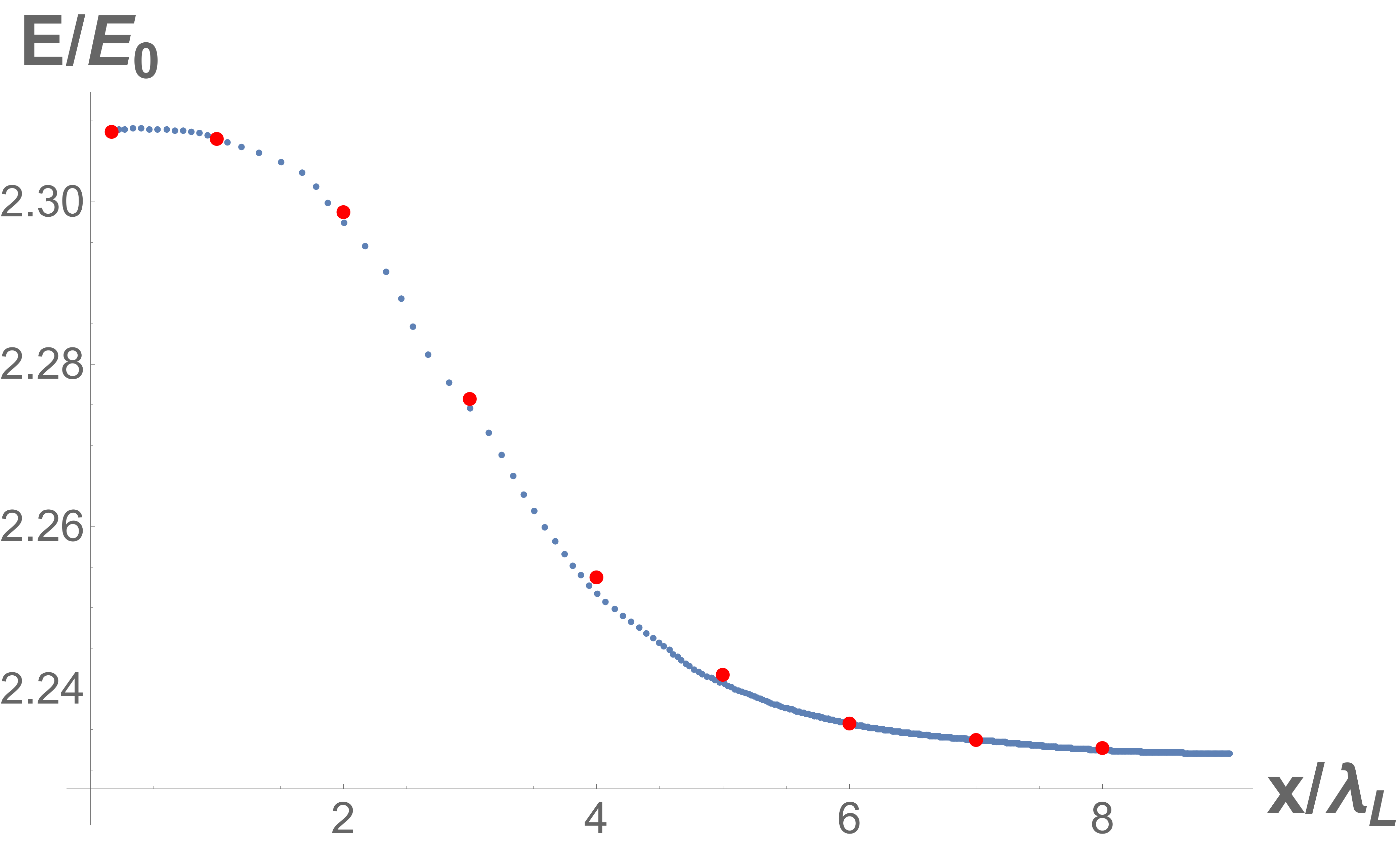}
    \caption{\small{Energy as a function of intervortex distance for $\kappa = 0.92$. The red dots are values taken from Ref.~\cite{jacobs}.}}
    \label{fig:endist}
\end{figure}
Another interesting issue concerns the modelling of large systems of nematic vortices in the presence of disorder. It is well known in many studies of dynamical phases of vortex matter, specially in cases where frozen disorder plays an important role, that vortices can be modelled as point particles subject to a pair-wise potential, the interaction with disorder and the influence of an external field \cite{daroca2010, Reichhardt2017}. To the best of our knowledge, this type of modelling has not been developed  for the case of nematic vortices. In this descriptions, where a set of rods is probably more adequate than point particles, the properties of the vortex-vortex interaction and the way it depends with distance and relative orientation plays a fundamental role. Our dynamical method allows to reconstruct and parametrize the involved force. Indeed, as by solving the TDGL we can easily obtain  the energy as a function of time $E(t)$, and the vortex-vortex separation as a functions of time $d(t)$, we can finally obtain $E(d)$ and from there the vortex-vortex force. We illustrate these ideas with an example for the standard GL theory (no nematicity). In  Fig.~\ref{fig:endist} our numerical results are compared with those obtained using variational methods \cite{jacobs}. This calculation can be easily implemented in the extended GL with nematicity, to then calculate the separation dependence of the vortex-vortex interactions. Furthermore, as we have access to the dynamics, we can compare the evolution under TDGL equations with the simpler rod model to improve the model or to bound errors. We mention too that relaxing the condition of $z-translation$ invariant solution is also trivial within the method. 

Finally, the method is specially suitable to study non equilibrium transport phenomena, a very relevant issue in order to compare with experimental results. Some of these ideas are part of work in progress or will be object of future work. Moreover, we are confident that, beside these problems, the method could be of general interest in the superconductivity community and beyond.

\section{Acknowledgements}

RSS, VB, GP and GSL acknowledge support by the University of Buenos Aires, UBACyT 20020170100496BA, Foncyt, PICT Raices -2019-2019-015890, PIP 11220150100653CO and CONICET. PDM acknowledges financial support from UBACYT 20020170100508BA and PICT Grant No. 2018-4298. EF acknowledges support by the US National Science Foundation under the grant DMR-1725401 at the University of Illinois.

\bibliographystyle{unsrt}

\bibliography{references}

\end{document}